\documentclass[twocolumn,reprint,superscriptaddress,prb,aps,showpacs,msmath,amsmath,amssymb0]{revtex4-2}
\usepackage{graphicx}
\usepackage{dcolumn}
\usepackage{bm}
\usepackage{subfigure}
\usepackage{natbib} 
\usepackage{multirow}
\usepackage{soul}
\usepackage{float}
\usepackage[normalem]{ulem}
\usepackage[usenames,dvipsnames]{color}
\usepackage{appendix}
\usepackage{mathrsfs}
\usepackage[colorlinks=true,linkcolor=blue,citecolor=blue,urlcolor=blue]{hyperref}

\begin{document}

\title{Understanding the role of Hubbard corrections in the rhombohedral phase of BaTiO$_3$}

\author{G. Gebreyesus}\email[]{ ghagoss@ug.edu.gh}
\affiliation{Department of Physics, School of Physical and Mathematical Sciences, College of Basic and Applied Sciences, University of Ghana, Accra, Ghana}

\author{Lorenzo Bastonero}
\affiliation{Bremen Center for Computational Materials Science, and MAPEX Center for Materials and Processes, University of Bremen, D-28359 Bremen, Germany}

\author{Michele Kotiuga}
\affiliation{Theory and Simulation of Materials (THEOS), and National Centre for Computational Design and Discovery of Novel Materials (MARVEL), \'Ecole Polytechnique F\'ed\'erale de Lausanne (EPFL), CH-1015 Lausanne, Switzerland}

\author{Nicola Marzari}
\affiliation{Bremen Center for Computational Materials Science, and MAPEX Center for Materials and Processes, University of Bremen, D-28359 Bremen, Germany}
\affiliation{Theory and Simulation of Materials (THEOS), and National Centre for Computational Design and Discovery of Novel Materials (MARVEL), \'Ecole Polytechnique F\'ed\'erale de Lausanne (EPFL), CH-1015 Lausanne, Switzerland}

\author{Iurii Timrov}\email[]{ iurii.timrov@psi.ch}
\affiliation{Theory and Simulation of Materials (THEOS), and National Centre for Computational Design and Discovery of Novel Materials (MARVEL), \'Ecole Polytechnique F\'ed\'erale de Lausanne (EPFL), CH-1015 Lausanne, Switzerland}
\affiliation{Present address: Laboratory for Materials Simulations (LMS), Paul Scherrer Institut (PSI), CH-5232 Villigen PSI, Switzerland}

\begin{abstract} 
We present a first-principles study of the low-temperature rhombohedral phase of BaTiO$_3$ using Hubbard-corrected density-functional theory. By employing density-functional perturbation theory, we compute the onsite Hubbard $U$ for Ti($3d$) states and the intersite Hubbard $V$ between Ti($3d$) and O($2p$) states. We show that applying the onsite Hubbard $U$ correction alone to Ti($3d$) states proves detrimental, as it suppresses the Ti($3d$)--O($2p$) hybridization and drives the system towards a cubic phase. Conversely, when both onsite $U$ and intersite $V$ are considered, the localized character of the Ti($3d$) states is maintained, while also preserving the Ti($3d$)--O($2p$) hybridization, restoring the rhombohedral phase of BaTiO$_3$. The generalized PBEsol+$U$+$V$ functional yields good agreement with experimental results for the band gap and dielectric constant, while the optimized geometry is slightly less accurate compared to PBEsol. Zone-center phonon frequencies and Raman spectra are found to be significantly influenced by the underlying geometry. PBEsol and PBEsol+$U$+$V$ provide satisfactory agreement with the experimental Raman spectrum when the PBEsol geometry is used, while PBEsol+$U$ Raman spectrum diverges strongly from experimental data highlighting the adverse impact of the $U$ correction alone in BaTiO$_3$. Our findings underscore the promise of the extended Hubbard PBEsol+$U$+$V$ functional with first-principles $U$ and $V$ for the investigation of other ferroelectric perovskites with mixed ionic-covalent interactions.
\end{abstract}

\date{\today} 

\maketitle

\section{Introduction}
\label{sec:intro}

BaTiO$_3$ (BTO) holds a prominent position among the extensively studied ABO$_3$ perovskite materials due to its wide range of technological applications in electronics, electromechanical energy conversion, nonlinear optics, and nonvolatile data storage~\cite{Peng-BOT-Appl-1, Rubio-Marcos2018-Appl-2, Sanna-2011, Waser-2003, de2015-appl-3, Mahadeva-7051095-appl-4, VASTA-20123071-appl-5}. It possesses a paraelectric cubic perovskite structure with no net polarization at high temperatures and undergoes a series of three ferroelectric phase transitions as the temperature decreases~\cite{kay-phase-trans, Merz-phase-trans}. Between 394~K and 278~K, BTO adopts a tetragonal structure with the polarization along the $\langle 100 \rangle$ crystal direction; as the temperature further decreases between 278~K and 183~K, it transforms into an orthorhombic structure with polarization along $\langle 110 \rangle$; finally, below 183~K, BTO adopts a rhombohedral structure with the $R3m$ space group with the polarization along $\langle 111 \rangle$~\cite{kay-phase-trans, Merz-phase-trans, Devonshire-phase-trans, Park-phase-trans}.  These phase transitions have been the subject of extensive experimental and theoretical investigations inspiring both the displacive~\cite{vonHippel:1946, Merz-phase-trans} and order-disorder~\cite{Bersuker:1966, Chaves:1976} models of ferroelectric transitions.

Computational studies based on density-functional theory (DFT)~\cite{Hohenberg:1964, Kohn-Sham-1965} have been instrumental in exploring the structural, electronic, optical, and vibrational properties of various phases of BTO. These studies have employed diverse exchange-correlation functionals, such as local-density approximation (LDA), generalized-gradient approximation (GGA), hybrid, and meta-GGA functionals ~\cite{Cohen, King, Wang, Evarestov_LDA_PBE_PBE0, GOH2016-BEC, Ghosez_BEC, Ghosez_LDA, hewat, Hermet_2009, Mahmoud-PBE0, Tsunoda_bandgap, Seo-BEC-Diel-B1WC, Zhang-BECs-Diele-2017, Mellaers-PBEsol-22, Yuk_PBEsol, Massimiliano, Michele-BOT-22}. Despite many results in good agreement with experiments, none of the aforementioned functionals provide an overall quantitative description of BTO. Consequently, the pursuit of even more accurate functionals has persisted. For instance, the PBEsol functional~\cite{Perdew:2008} has shown high accuracy in predicting the structure of the rhombohedral phase of BTO, however, it underestimates the band gap~\cite{Mellaers-PBEsol-22, Yuk_PBEsol}. On the other hand, hybrid functionals like PBE0~\cite{Adamo:1999} and HSE06~\cite{Heyd:2003, Heyd:2006} improve the band-gap prediction, but they overestimate the lattice constant and the atomic distortions associated with ferroelectricity~\cite{Evarestov_LDA_PBE_PBE0, Tsunoda_bandgap}. As a result, any physical property dependent on atomic distortions (e.g., phonons) is significantly influenced by the choice of the functional~\cite{PhysRevMaterials-Moseni-phonon-polar}. In turn, this affects the accuracy of vibrational properties, leading to shifts in computed Raman spectral peaks or incorrect intensities as compared to experiments. Hence, there is a pressing need to search for novel functionals capable of providing an accurate characterization of the structural, electronic and vibrational properties of BTO simultaneously. 

One of the widely used approaches to model transition-metal oxides is DFT+$U$~\cite{anisimov:1991, Anisimov-1997-DFT+U, dudarev1998electron}, where the Hubbard $U$ correction is applied to selected states (typically to partially filled $d$ states) to alleviate self-interaction errors~\cite{Kulik:2006, Kulik:2008, Cohen:2008}. The application of DFT+$U$ to BTO has only gained prominence in the last decade~\cite{Zhang:2014, Erhart:2014, Liu:2017, Chen:2017, Watanabe:2018, Majumder:2019, Tsunoda_bandgap, Din:2020, Dwij:2022}. The reluctance to apply this approach with the Hubbard $U$ correction on Ti($3d$) states in BTO in early DFT-based works was primarily due to the fact that Ti ions are in the $4+$ oxidation state (OS), resulting in a $d^0$ electronic configuration. Thus, in BTO the $3d$ states of Ti are nominally empty, suggesting that the application of Hubbard $U$ would have a negligible effect. Conversely, in BTO with oxygen vacancies or in BTO oxyhydrides, some Ti ions undergo reduction to $3+$ OS with a $d^1$ configuration, which provides a stronger motivation for the use of Hubbard corrections~\cite{Zhang:2014, Liu:2017, Majumder:2019}.  However, earlier DFT studies of the pristine BTO employing LDA and GGA revealed a significant hybridization between Ti($3d$) and O($2p$) states, indicating that the four electrons of the Ti($3d$) states are not entirely transferred to the neighboring O($2p$) states (see, e.g., Ref.~\cite{Ghosez_LDA}). In fact, this hybridization plays a crucial role in the covalency of the Ti-O bonds, which is essential for the manifestation of ferroelectricity~\cite{Cohen, Cohen_nature, KATAR, Ghosez_LDA, Resta:1993}. Therefore, the effect of the $U$ correction on Ti($3d$) states should not be presumed to be exactly zero. In fact, applying the Hubbard $U$ correction to the Ti($3d$) states localizes them more on Ti ions and erroneously suppresses hybridizations with the O($2p$) states (see Fig.~5 in Ref.~\cite{Din:2020}). Consequently, the rhombohedral distortion disappears, and instead, the cubic phase of BTO becomes more stable (see Fig.~S6 in Ref.~\cite{Tsunoda_bandgap}). This unequivocally demonstrates that the effect of onsite $U$ corrections on Ti($3d$) states is detrimental in pristine BTO and underscores the crucial importance of preserving hybridizations with O($2p$) states. In this regard, the extended DFT+$U$+$V$ formulation~\cite{campo2010extended} proves highly attractive as it allows for the application of an onsite $U$ correction to the Ti($3d$) states to reduce self-interaction errors while ensuring the intersite Ti($3d$)--O($2p$) hybridizations thanks to the $V$ corrections, thereby preserving the fundamental covalency of the Ti-O bonds essential for ferroelectricity in BTO~\cite{Cohen_nature}.

The main challenge of the DFT+$U$ approach is the lack of \textit{a priori} knowledge of the value for $U$. Even though there are various first-principles methods to compute it~\cite{Himmetoglu:2014}, $U$ is still often determined empirically. Despite numerous DFT+$U$ studies of BTO~\cite{Zhang:2014, Erhart:2014, Liu:2017, Chen:2017, Watanabe:2018, Majumder:2019, Tsunoda_bandgap, Din:2020, Dwij:2022}, no consensus has been reached regarding which states should undergo the Hubbard correction and which value of $U$ should be used. Additionally, differences between various types of Hubbard projector functions are often overlooked when comparing the $U$ values from different DFT+$U$ studies of various materials~\cite{Wang:2016, Kick:2019}. For example, in previous DFT+$U$ investigations of BTO, Ti($3d$) states were corrected using the Hubbard $U$ value of 3~eV~\cite{Majumder:2019} or 4.49~eV~\cite{Zhang:2014, Tsunoda_bandgap, Liu:2017}, while in Refs.~\cite{Erhart:2014, Watanabe:2018} the Hubbard $U$ correction was applied to O($2p$) states with the value of 8~eV. Furthermore, in Ref.~\cite{Dwij:2022} the $U$ correction of 8~eV was applied to both Ti($3d$) and O($2p$) states. Additionally, the choice of Hubbard projector functions also varied in these works: Most studies employed projector-augmented wave (PAW) Hubbard projector functions, except for Ref.~\cite{Zhang:2014} where a different type of projector was utilized (not reported in that paper, but either nonorthogonalized or orthogonalized atomic orbitals~\cite{Timrov:2020b}). Moreover, the value of $U$ was determined empirically in all the aforementioned DFT+$U$ studies, with only one exception~\cite{Zhang:2014}, where it was computed using linear-response theory~\cite{cococcioni2005linear}. As a result of this large variation in $U$ values and the ambiguity surrounding the correction of Ti($3d$) or O($2p$) states or both, there exists a wide spread of results that often contradict both each other and experimental observations. Furthermore, as of now, no DFT+$U$+$V$ studies of BTO have been conducted to investigate the significance of intersite Hubbard $V$ corrections between Ti($3d$) and O($2p$) states.

In this study, we present a comprehensive first-principles investigation of the low-temperature rhombohedral phase of BTO, focusing on its structural, electronic, and vibrational properties using three functionals: PBEsol, PBEsol+$U$, and PBEsol+$U$+$V$. To determine the onsite $U$ and intersite $V$ Hubbard parameters, we employ a rigorous first-principles approach based on linear-response theory~\cite{cococcioni2005linear}, recast in terms of density-functional perturbation theory (DFPT)~\cite{timrov2018hubbard, Timrov:2021} in a basis of L\"owdin-orthogonalized atomic orbitals (Hubbard projector functions). This approach eliminates any empirical input and potential ambiguities typically present in Hubbard-corrected DFT studies. Furthermore, the self-consistent procedure is employed for computing the Hubbard parameters~\cite{Timrov:2021} to ensure the mutual consistency of the crystal and electronic structures. In agreement with Ref.~\cite{Tsunoda_bandgap}, we find that applying the onsite $U$ correction solely to Ti($3d$) states results in a pronounced suppression of hybridization with neighboring O($2p$) states, driving the system into a cubic phase. However, our DFT+$U$ calculations, utilizing first-principles $U$ value, provide the cubic structure that is still exhibiting dynamical instability with imaginary phonon modes around the $\Gamma$ point, in contrast to the results of Ref.~\cite{Tsunoda_bandgap}. Conversely, the introduction of intersite Hubbard $V$ interactions between the Ti($3d$) and O($2p$) states dramatically alters the overall picture by restabilizing the rhombohedral phase thanks to the restored covalency of the Ti-O bonds. The optimized geometry obtained from PBEsol+$U$+$V$ is found to be somewhat less accurate than that from PBEsol, whereas the projected density of states (PDOS) is qualitatively very similar in both cases. In addition, the Born effective charges (BEC) are slightly smaller for certain components within PBEsol+$U$+$V$ as compared to PBEsol, while the dielectric constant and band gap from PBEsol+$U$+$V$ exhibit good agreement with experimental data, surpassing the accuracy of PBEsol predictions by a significant margin. On the contrary, all predictions obtained from PBEsol+$U$ are systematically less accurate compared to the PBEsol+$U$+$V$ case. Lastly, the zone-center phonon frequencies and Raman spectra are found to be highly sensitive to the underlying geometry. The PBEsol and PBEsol+$U$+$V$ Raman spectra are found to be in satisfactory agreement with experiments provided that the PBEsol geometry is used, while the Raman spectrum from PBEsol+$U$ differs dramatically from the experimental one.
	
The remainder of this paper is organized as follows: Section~\ref{sec:comput_details} presents the computational details; Sec.~\ref{sec:Results_and_Discussion} contains an in-depth analysis of the results, encompassing the structural properties, PDOS, BEC, the dielectric tensor, phonon dispersions, and Raman spectra; and finally, Sec.~\ref{sec:conclusions} contains the concluding remarks.

\section{Computational details}
\label{sec:comput_details}

All calculations are performed using the plane-wave pseudopotential method as implemented in the \textsc{Quantum ESPRESSO} distribution~\cite{Giannozzi:2009, Giannozzi:2017, Giannozzi:2020}. We use the exchange-correlation functional constructed using GGA with the PBEsol parametrization~\cite{Perdew:2008}, and we employ the following pseudopotentials from the Pslibrary v1.0.0~\cite{DALCORSO2014337}: Ba.pbesol-spn-rrkjus\_psl.1.0.0.UPF, Ti.pbesol-spn-rrkjus\_psl.1.0.0.UPF, and O.pbesol-n-rrkjus\_psl.1.0.0.UPF. The Kohn-Sham wavefunctions and potentials are expanded in plane waves up to a kinetic-energy cutoff of 80 and 800~Ry, respectively. The first Brillouin zone is sampled using uniform $\mathbf{k}$-point meshes of size $8 \times 8 \times 8$ and $18 \times 18 \times 18$  centered at the $\Gamma$ point for the ground state and PDOS calculations, respectively. PDOS is obtained using the tetrahedron method~\cite{Blochl:1994}.

\begin{figure}[t]
	\centering
	\includegraphics[width=0.98\linewidth]{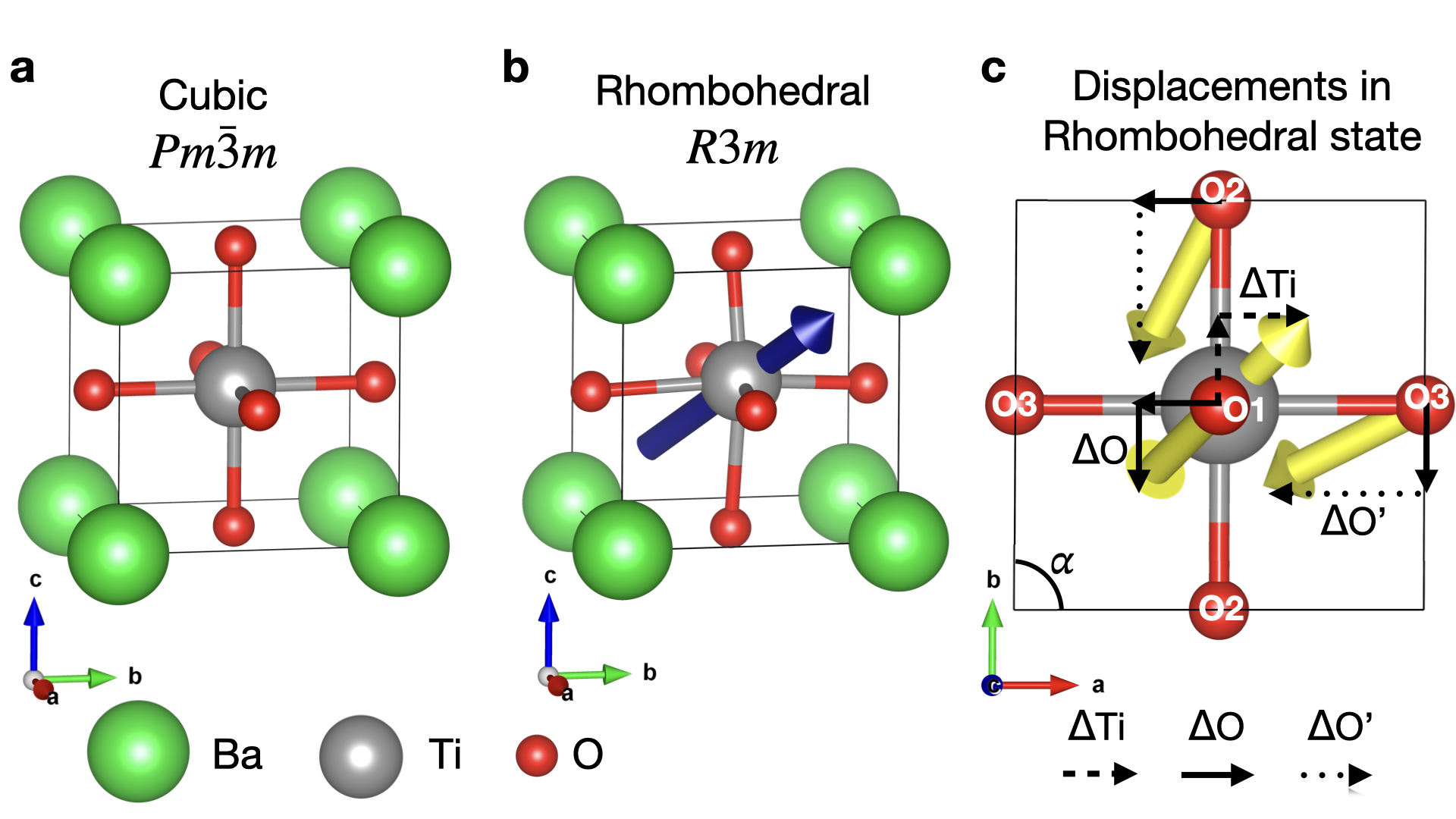}
	\caption{Crystal structure of BTO in the (a)~cubic and (b)~rhombohedral perovskite structures with the Ba, Ti, and O atoms represented in green, gray, and red, respectively. While the cubic structure is nonpolar, the rhombohedral phase exhibits a polarization along the $\langle 111 \rangle$ direction, denoted by the dark blue arrow. (c)~Schematic illustration of the Ti and O atomic displacements in the rhombohedral structure relative to the cubic one projected onto the $ab$ plane. The total displacements are depicted with yellow arrows, while the components relative to the cubic structure are indicated with black arrows: Dashed lines for $\Delta_{\mathrm{Ti}}$ and solid and dotted lines for the two O displacements $\Delta_{\mathrm{O}}$ and $\Delta_{\mathrm{O'}}$. For clarity, Ba atoms are omitted. The rotational symmetry around the $\langle 111 \rangle$ direction results in identical displacements when viewed in the $ac$ and $bc$ planes.}
	\label{fig:cryst_structure}
\end{figure}

The low-temperature rhombohedral phase of BTO is described by its lattice constant, rhombohedral angle, and symmetry-preserving internal atomic displacements along the $\langle 111 \rangle$ direction. The atomic positions of the rhombohedral crystal structure can be expressed based on those in the cubic structure (in crystal coordinates) as follows: 
\begin{center}
	\begin{tabular}{cc}
		Ba:& (0.0 + $\Delta_\mathrm{Ba}$, 0.0 + $\Delta_\mathrm{Ba}$, 0.0 + $\Delta_\mathrm{Ba}$),\\
		Ti:& (0.5 + $\Delta_\mathrm{Ti}$, 0.5 + $\Delta_\mathrm{Ti}$, 0.5 + $\Delta_\mathrm{Ti}$),\\
		O1:& (0.5 + $\Delta_\mathrm{O}$, 0.5 + $\Delta_\mathrm{O}$, 0.0 + $\Delta_\mathrm{O'}$),\\
		O2:& (0.5 + $\Delta_\mathrm{O}$, 0.0 + $\Delta_\mathrm{O'}$, 0.5 + $\Delta_\mathrm{O}$),\\
		O3:& (0.0 + $\Delta_\mathrm{O'}$, 0.5 + $\Delta_\mathrm{O}$, 0.5 + $\Delta_\mathrm{O}$),
	\end{tabular}
\end{center}
where $\Delta_\mathrm{Ba}$, $\Delta_\mathrm{Ti}$, $\Delta_\mathrm{O}$, and $\Delta_\mathrm{O'}$ represent the displacements of the Ba, Ti, and O ions from their atomic positions in the cubic phase (see Fig.~\ref{fig:cryst_structure}). We perform a full structural relaxation to calculate these displacements using three functionals (PBEsol, PBEsol+$U$, and PBEsol+$U$+$V$). The Broyden-Fletcher-Goldfarb-Shanno (BFGS) algorithm~\cite{Fletcher:1987} is employed for geometry optimization with convergence thresholds set to $10^{-8}$~Ry, $10^{-5}$~Ry/bohr, and $0.01$~Kbar for the total energy, forces, and pressure, respectively.

For the PBEsol+$U$ and PBEsol+$U$+$V$ calculations, we determine the onsite $U$ and intersite $V$ Hubbard parameters self-consistently using DFPT~\cite{timrov2018hubbard, Timrov:2021}, implemented in the \textsc{HP} code~\cite{TIMROV-HP-2022} of \textsc{Quantum ESPRESSO}~\cite{Giannozzi:2009, Giannozzi:2017, Giannozzi:2020}. The reader is invited to check Refs.~\cite{timrov2018hubbard, Timrov:2021} for the detailed explanations on how the Hubbard parameters are defined and computed using DFPT. As Hubbard projector functions, we employ the L\"owdin-orthogonalized atomic orbitals~\cite{Lowdin:1950, Mayer:2002, Timrov:2020b}. We use uniform $\Gamma$-centered $\mathbf{k}$- and $\mathbf{q}$-point meshes of size $6\times 6\times 6$ and $4 \times 4 \times 4$, respectively, in computing the Hubbard parameters with an accuracy of $\sim 0.01$~eV for $U$ and $\sim 0.001$~eV for $V$. Employing a self-consistent procedure~\cite{Timrov:2021}, our calculations encompass cyclic iterations that involve structural optimizations and successive recalculations of Hubbard parameters for each new geometry. By doing so, we obtain structural parameters that are optimized using PBEsol+$U$ and PBEsol+$U$+$V$ using the respective self-consistent Hubbard parameters, and that are listed in Table~\ref{table:structural-optimization}. This methodology has demonstrated its efficacy in yielding accurate results for various transition-metal compounds~\cite{Timrov:2020c, Mahajan:2021, Timrov:2022c, Mahajan:2022, Timrov:2023, Haddadi:2023, Binci:2023, Bonfa:2023}. The computed values of the Hubbard parameters are as follows: $U=4.52$~eV for Ti($3d$) states within PBEsol+$U$, and $U=5.21$~eV for Ti($3d$) states with intersite $V$ values ranging from 1.17 to 1.37~eV (depending on the interatomic distance) between Ti($3d$) and O($2p$) states within PBEsol+$U$+$V$. These values are consistent with previous studies of SrTiO$_3$ using the same approach~\cite{Ricca:2020}. 

It is important to note that Hubbard parameters are not universal~\cite{timrov2018hubbard, Timrov:2021, TIMROV-HP-2022}: Their values depend on the chemical environment of transition-metal elements, oxidation state~\cite{Kulik:2008}, spin configuration~\cite{Mahajan:2021, Mahajan:2022}, and other factors. When it comes to the material phase's influence on computed Hubbard parameters, variations can arise, especially if distinct phases trigger notable changes in the chemical environment of transition-metal elements. For instance, in the case of the tetragonal phase of BTO, we obtain the following values: $U=4.50$~eV for Ti($3d$) states within PBEsol+$U$, and $U=5.25$~eV for Ti($3d$) states with intersite $V$ values ranging from 1.29 to 1.32~eV between Ti($3d$) and O($2p$) states within PBEsol+$U$+$V$. As can be seen, the differences between the Hubbard parameters of tetragonal and rhombohedral phases are very small ($<1\%$), which is not surprising since the crystallographic differences between these two phases are small. Consequently, our computed Hubbard parameters for the rhombohedral phase are transferrable to other phases of BTO, provided identical computational setups, like pseudopotentials and Hubbard projector functions, are used. They can, at the very least, serve as very good starting point when refining these parameters for other phases using DFPT. However, in general it is important to recompute Hubbard parameters for different phases of the same material and verify how significant are the changes.

To compute the phonon frequencies, we utilize the frozen-phonon method implemented in the \textsc{Phonopy} package~\cite{togo_first_2015}, using atomic displacements of 0.01~\AA. The $3 \times 3 \times 3$ supercell is employed to calculate the interatomic force constants. We verified that adopting larger supercells of size $4 \times 4 \times 4$ does not significantly alter the phonon dispersions. The first Brillouin zone is sampled using a uniform $4 \times 4 \times 4$ $\mathbf{k}$-point mesh centered at the $\Gamma$ point. To correct for the nonanalyticity of the dynamical matrix as $\mathbf{q}\rightarrow \mathbf{0}$ we include up to dipole-dipole interactions which depend on the BEC and the dielectric tensor~\cite{baroni_phonons_2001}. We note that Royo et al.~\cite{Massimiliano} investigated the effect of higher-order corrections, which we discuss but ultimately neglect in the results presented here. For the calculation of the BEC and the dielectric tensor, we employ two approaches: the finite electric field method~\cite{umari_ab_2002, Souza:2002, Umari2005a} and DFPT~\cite{giannozzi_ab_1991, gonze_dynamical_1997, baroni_phonons_2001, tobik_electric_2004}. The method of Refs.~\cite{umari_ab_2002, Souza:2002, Umari2005a} for computing BEC, dielectric, Raman and nonlinear optical susceptibility tensors is used as implemented in the \texttt{aiida-vibroscopy}~\cite{Bastonero2023} package, exploiting AiiDA~\cite{Huber2020, Uhrin2021}. The directional-sampling technique described in Ref.~\cite{Bastonero2023} is used to effectively sample the Brillouin zone with two sets of $\mathbf{k}$-point distances, namely the ``parallel'' and the ``orthogonal'' (referring to the direction of the applied electric field). For the parallel and orthogonal distances, we set values of 0.1~\AA\ and 0.3~\AA, respectively (we note that 0.1~\AA\ corresponds to the $16\times16\times16$ uniform $\mathbf{k}$-point mesh). The BEC and the dielectric tensor are defined in the real-space rhombohedral BTO reference system with the $z$ axis parallel to the $C_3$ rhombohedral axis (i.e., the $\langle 111 \rangle$ direction), while the $x$ axis is perpendicular to it. The zone-center phonons and the phonon dispersions are computed for phonon wavevectors $\mathbf{q}$ defined in reciprocal space of the rhombohedral BTO reference system.

To calculate the nonresonant Raman intensities, we adopt the Placzek approximation (first-order processes)~\cite{Placzek-Lee-1983}. In order to compare with experiments from Refs.~\cite{Tenne-Ramman, Maslova2020}, for Raman spectrum calculations we use the real-space tetragonal BTO reference system with the $z$ axis being parallel to the $c$ tetragonal axis, and the $x$ axis is perpendicular to it. Within this framework, we consider the backscattering geometry with a parallel polarization configuration, namely the $z(xx)\bar{z}$ setup in Porto notation~\cite{Damen:1966}. This entails that the incident light's \textit{propagation direction} aligns with $z$, while the scattered light follows $\bar{z}$. Both the incident and scattered light's \textit{polarization direction} align with $x$. The Raman scattering amplitude for a phonon mode $\nu$ is expressed as~\cite{Brueesch1986}
\begin{equation}
    I_{\nu} 
    \propto
    (\omega_L-\omega_{\nu})
    \frac{n_{\nu}+1}{\omega_{\nu}}
    |
    \mathbf{e}_i \cdot
    \overset{\leftrightarrow}{\mathbf{A}}_{\nu}
    \cdot \mathbf{e}_s
    |^2
    ~,
    \label{eq:ramanamplitude}
\end{equation}
where $n_{\nu}$ represents the Bose-Einstein occupation, $\overset{\leftrightarrow}{\mathbf{A}}_{\nu}$ denotes the Raman susceptibility tensor, $\omega_L$ is the incident laser frequency, $\omega_{\nu}$ denotes the frequency of the phonon mode $\nu$, and $\mathbf{e}_i$ and $\mathbf{e}_s$ are the polarization vectors of the incident and scattered light, respectively. The temperature relevant for the Bose-Einstein occupation and the laser frequency are adjusted to match the experimental conditions. For the $z(xx)\bar{z}$ setup, $\mathbf{e}_i$ and $\mathbf{e}_s$ are parallel to each other, both lying along the $x$ direction. While the propagation directions of the incident ($\mathbf{k}_i$) and scattered ($\mathbf{k}_s$) light do not appear explicitly in Eq.~\eqref{eq:ramanamplitude}, their difference $\mathbf{q} = \mathbf{k}_s-\mathbf{k}_i$ is utilized to specify the direction for the limit $\mathbf{q} \rightarrow \mathbf{0}$ when computing the nonanalytic component of the dynamical matrix and the Raman susceptibility tensor~\cite{Popov:2020}. 
To model the experimental backscattering geometry, $\mathbf{q}$ is chosen to be parallel to the $z$ axis in reciprocal space of the tetragonal BTO reference system, as the photon propagates back and forth along this direction. To generate the Raman spectra, we employ a Lorentzian smearing function with a constant broadening parameter of 8 cm$^{-1}$.

The data used to generate the results presented in this paper are accessible in the Materials Cloud Archive~\cite{MaterialsCloudArchive2023}.
 
\section{Results and Discussion}
\label{sec:Results_and_Discussion}

\subsection{Structural properties}
\label{structural}

In this section, we present the results of structural optimizations for the low-temperature rhombohedral phase of BTO using PBEsol, PBEsol+$U$, and PBEsol+$U$+$V$ functionals. A summary of our findings and a comparison with previous computational studies based on DFT with (semi-)local and hybrid functionals~\cite{Evarestov_LDA_PBE_PBE0, Yuk_PBEsol, Huai-LDA, Seo2013}, as well as experimental data~\cite{kwei}, are presented in Table~\ref{table:structural-optimization}. It is well known that LDA underestimates the lattice parameters~\cite{Huai-LDA}, while PBE tends to overestimate them~\cite{Evarestov_LDA_PBE_PBE0}, which is also observed in the case of BTO. Our PBEsol results are consistent with previous PBEsol studies~\cite{Yuk_PBEsol}, showing very good agreement with the experimental lattice parameter $a$, rhombohedral angle $\alpha$, and atomic displacements. On the other hand, previous DFT studies using PBE0~\cite{Yuk_PBEsol} and B1-WC~\cite{Seo2013} hybrid functionals either significantly overestimate or underestimate $a$, respectively, while $\alpha$ is closer to the experimental value. As mentioned in Sec.~\ref{sec:intro}, the PBEsol+$U$ and PBEsol+$U$+$V$ predictions for structural properties are in stark contrast to each other: The former drives the structure towards the cubic phase, while the latter preserves the rhombohedral phase (see Table~\ref{table:structural-optimization}). The PBEsol+$U$ and PBEsol+$U$+$V$ structural parameters are discussed in more detail in the following.

\begin{center}
	\begin{table}[t]
    \renewcommand{\arraystretch}{1.3}
	\begin{tabular}{lccccc} 
        \hline\hline
		& \parbox{1.1cm}{$a$ (\AA)} & \parbox{1.2cm}{$\alpha$ (deg)} & \parbox{1.1cm}{$\Delta_\mathrm{Ti}$} & \parbox{1.1cm}{$\Delta_\mathrm{O}$} & \parbox{1.1cm}{$\Delta_\mathrm{O'}$} \\ \hline
		PBEsol                             & 3.998 & 89.87 & $-0.012$           & 0.011 & 0.018 \\
		PBEsol+$U$                         & 3.990 & 90.00 & $\phantom{-}0.000$ & 0.000 & 0.000 \\
		PBEsol+$U$+$V$                     & 4.017 & 89.77 & $-0.014$           & 0.012 & 0.020 \\ 
        LDA~\cite{Huai-LDA}                & 3.931 & 89.92 & $-0.007$           & 0.010 & 0.014 \\  
        PBE~\cite{Evarestov_LDA_PBE_PBE0}  & 4.073 & 89.71 & $-0.015$           & 0.014 & 0.025 \\  
		PBEsol~\cite{Yuk_PBEsol}           & 3.998 &       & $-0.012$           & 0.012 & 0.019 \\ 
        PBE0~\cite{Evarestov_LDA_PBE_PBE0} & 4.029 & 89.73 & $-0.015$           & 0.013 & 0.024 \\ 
        B1-WC~\cite{Seo2013}               & 3.991 & 89.86 &                    &       &       \\
		Expt.~\cite{kwei}                  & 4.004 & 89.84 & $-0.013$           & 0.011 & 0.019 \\ 
       \hline\hline
	\end{tabular}	
    \caption{Comparison of the optimized lattice parameter $a$, the rhombohedral angle $\alpha$, and the atomic displacements (in crystal coordinates) $\Delta_\mathrm{Ti}$, $\Delta_\mathrm{O}$, and $\Delta_\mathrm{O'}$ (see Fig.~\ref{fig:cryst_structure}) computed in this work using PBEsol, PBEsol+$U$, and PBEsol+$U$+$V$ and as obtained in previous computational studies~\cite{Evarestov_LDA_PBE_PBE0, Yuk_PBEsol, Huai-LDA, Seo2013} and in experiments at 15~K~\cite{kwei}. $\Delta_\mathrm{Ba} = 0$ in all cases. The Hubbard $U$ correction is applied only to the Ti($3d$) states while $V$ is applied between Ti($3d$) and O($2p$).}
	\label{table:structural-optimization}
\end{table}
\end{center}

Let us compare the accuracy of the structural parameters obtained using PBEsol and PBEsol+$U$+$V$. As can be seen in Table~\ref{table:structural-optimization}, PBEsol underestimates $a$ by only $\sim 0.2\%$, and $\alpha$ is overestimated by only $\sim 0.03\%$. These results are significantly better than those obtained with LDA and PBE~\cite{Huai-LDA, Evarestov_LDA_PBE_PBE0}. On the other hand, PBEsol+$U$+$V$ overestimates $a$ by $\sim 0.3\%$ and underestimates $\alpha$ by $\sim 0.08\%$. Such seemingly minor deterioration in the accuracy of structural predictions within PBEsol+$U$+$V$ compared to PBEsol has a significant impact on the lattice vibrational properties, as evidenced in Secs.~\ref{phonon-dispersion} and \ref{Raman}. 

Motivated by Refs.~\cite{Erhart:2014, Watanabe:2018}, we examine the impact of applying the Hubbard $U$ correction to O($2p$) states. First, we determine $U$ for the O($2p$) states from first principles using DFPT in a self-consistent fashion, and we find the values of 8.6 and 8.7~eV within PBEsol+$U$ and PBEsol+$U$+$V$, respectively. Applying the $U$ correction to both O($2p$) and Ti($3d$) states within PBEsol+$U$ and PBEsol+$U$+$V$ results in the optimized geometry of rhombohedral BTO adopting a cubic structure. This occurs as $U$ localizes O($2p$) electrons, diminishes Ti--O hybridization, suppressing covalency, and ultimately stabilizing the cubic phase. However, this contradicts experimental observations where the rhombohedral phase prevails. Consequently, we abstain from computing additional properties such as electronic structure and vibrational spectra that involve the $U$ correction on O($2p$). In the rest of the paper we neglect the effect of $U$ on O($2p$) states.

\begin{figure}[t]
	\centering
	\includegraphics[width=0.49\textwidth]{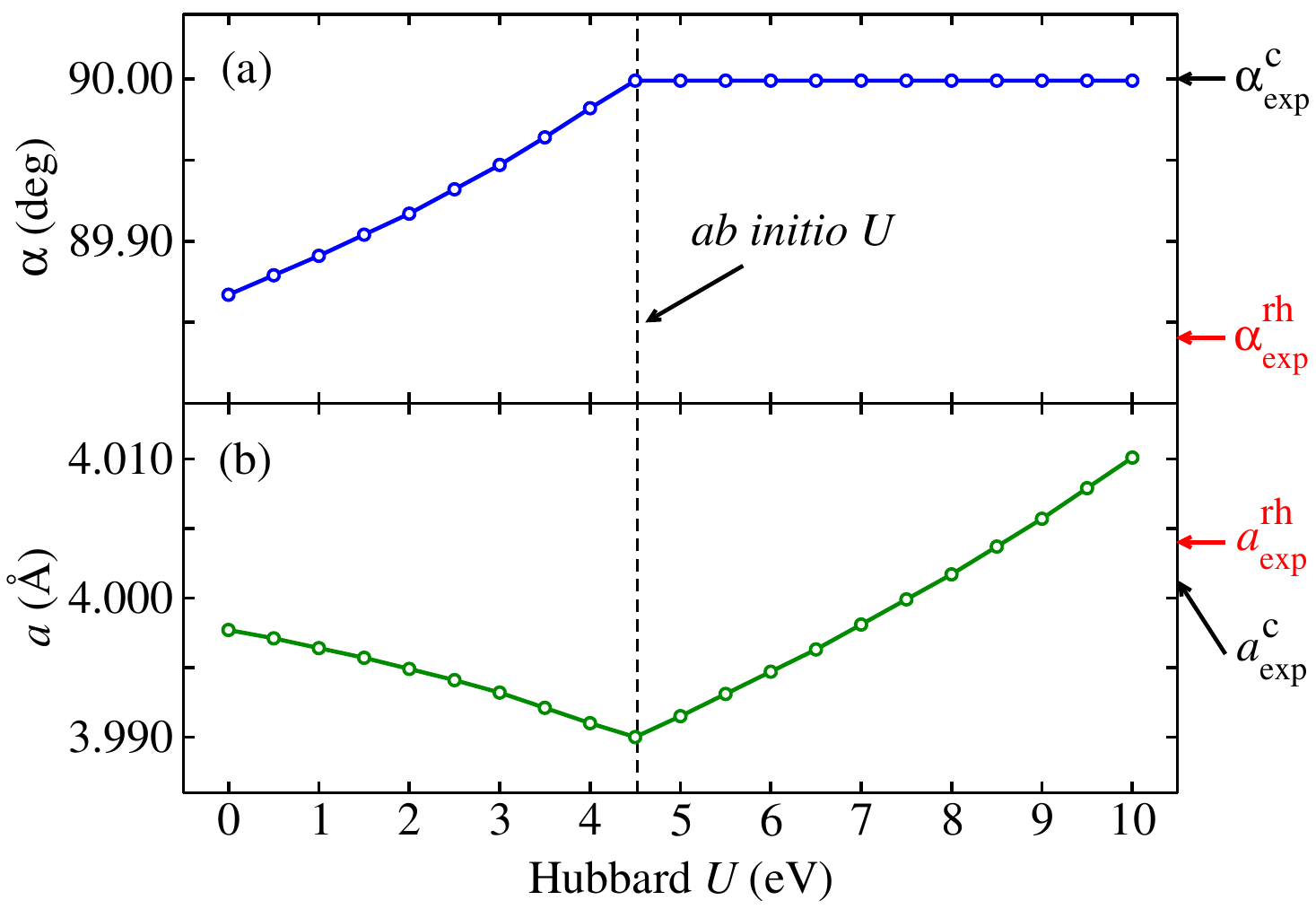}
	\caption{Optimized lattice parameter $a$ (a) and the rhombohedral angle $\alpha$ (b) as a function of the Hubbard $U$ parameter applied to Ti($3d$) states using PBEsol+$U$. The \textit{ab initio} value of $U$ is 4.52~eV and it was computed using DFPT, as discussed in Sec.~\ref{sec:comput_details}. Experimental values of the structural parameters are also indicated: for the rhombohedral phase these are $a_\mathrm{exp}^\mathrm{rh} = 4.004$~\AA\ and $\alpha_\mathrm{exp}^\mathrm{rh} = 89.84^\circ$ at 15~K~\cite{kwei}, and for the cubic phase these are $a_\mathrm{exp}^\mathrm{c} = 4.001$~\AA\ and $\alpha_\mathrm{exp}^\mathrm{c} = 90.00^\circ$ at 400~K~\cite{Megaw:1947}.}
	\label{fig:Alpha_vs_U}
\end{figure}

The stabilization of the cubic phase of BTO within PBEsol+$U$ requires further analysis. To investigate this, we performed structural optimizations starting from the experimental rhombohedral structure while varying the value of $U$ in the range from 0 to 10~eV. Figure~\ref{fig:Alpha_vs_U} illustrates the variation of lattice parameter $a$ and rhombohedral angle $\alpha$ as a function of $U$. Increasing $U$ from 0 to 4.5~eV leads to decreasing of $a$ in a quasilinear fashion by $\sim 0.2\%$ and a quasilinear increase in $\alpha$ by $\sim 0.1\%$, leading it towards $90^\circ$. Further increasing $U$ beyond 4.5~eV results in a larger $a$ (also quasilinear change), while $\alpha$ remains stable at $90^\circ$. Hence, a critical value of $U$ is observed at approximately 4.5~eV, where the cubic phase becomes stable. Interestingly, our first-principles value of $U$ is found to be 4.52~eV, which aligns well with this critical value. It is important to note that these $U$ values were determined using L\"owdin-orthogonalized atomic orbitals as Hubbard projectors (see Sec.~\ref{sec:comput_details}), and hence they may not be directly applicable with other types of Hubbard projectors, necessitating reevaluation for such cases. Therefore, we find that by increasing $U$ from 0 to 4.5~eV the unit cell volume is decreasing and the rhombohedral distortion smoothly vanishes, while when further increasing $U$ the cubic structure remains stable and its volume increases. Why is there such a nonmonotonic behavior of the cell volume as a function of $U$? The increase in $U$ results in a more localized nature of Ti($3d$) states and a reduction in their hybridization with the neighboring O($2p$) states, consequently diminishing the covalent character of Ti($3d$)--O($2p$) bonds. This, in turn, causes the Ti and O ions to occupy high-symmetry positions of the cubic phase. As a result of such an interplay between structural and electronic degrees of freedom, overall the unit cell volume tends to decrease. However, when $U$ is increased beyond the critical value, a typical behavior is observed: Larger $U$ values lead to even stronger localization of Ti($3d$) states and a more pronounced ionic character of interactions, which consequently expands the lattice (increase in $a$).

As mentioned earlier, the inclusion of intersite Hubbard $V$ interactions between Ti($3d$) and O($2p$) states plays a crucial role in preserving the hybridizations between these states and, consequently, in maintaining the rhombohedral distortion of the lattice. As indicated in Sec.~\ref{sec:comput_details}, within PBEsol+$U$+$V$, the $U$ value on Ti($3d$) states is found to be 5.21~eV, which is higher than the value obtained within PBEsol+$U$. This behavior is common and is related to changes in electronic screening when computing both $U$ and $V$~\cite{Timrov:2021, Mahajan:2021, Mahajan:2022}. Despite the fact that the computed $U$ value within PBEsol+$U$+$V$ exceeds the critical $U$ value obtained within PBEsol+$U$, the cubic phase does not appear. This observation underscores the strong impact of intersite $V$ interactions in stabilizing the rhombohedral phase, even though the $V$ values are much smaller than the $U$ values.

In the next section, we utilize the optimized structural parameters obtained for each of the considered functionals (if not otherwise stated) to analyze the effects of Hubbard corrections on the electronic and vibrational properties of BTO. This analysis provides further insights into the role of extended Hubbard functionals with both $U$ and $V$ interactions in accurately predicting various properties of the rhombohedral phase of BTO.

\subsection{Projected density of states and band gap}
\label{projected-dos}

\begin{figure}[b]
	\centering
	\includegraphics[width=0.47\textwidth]{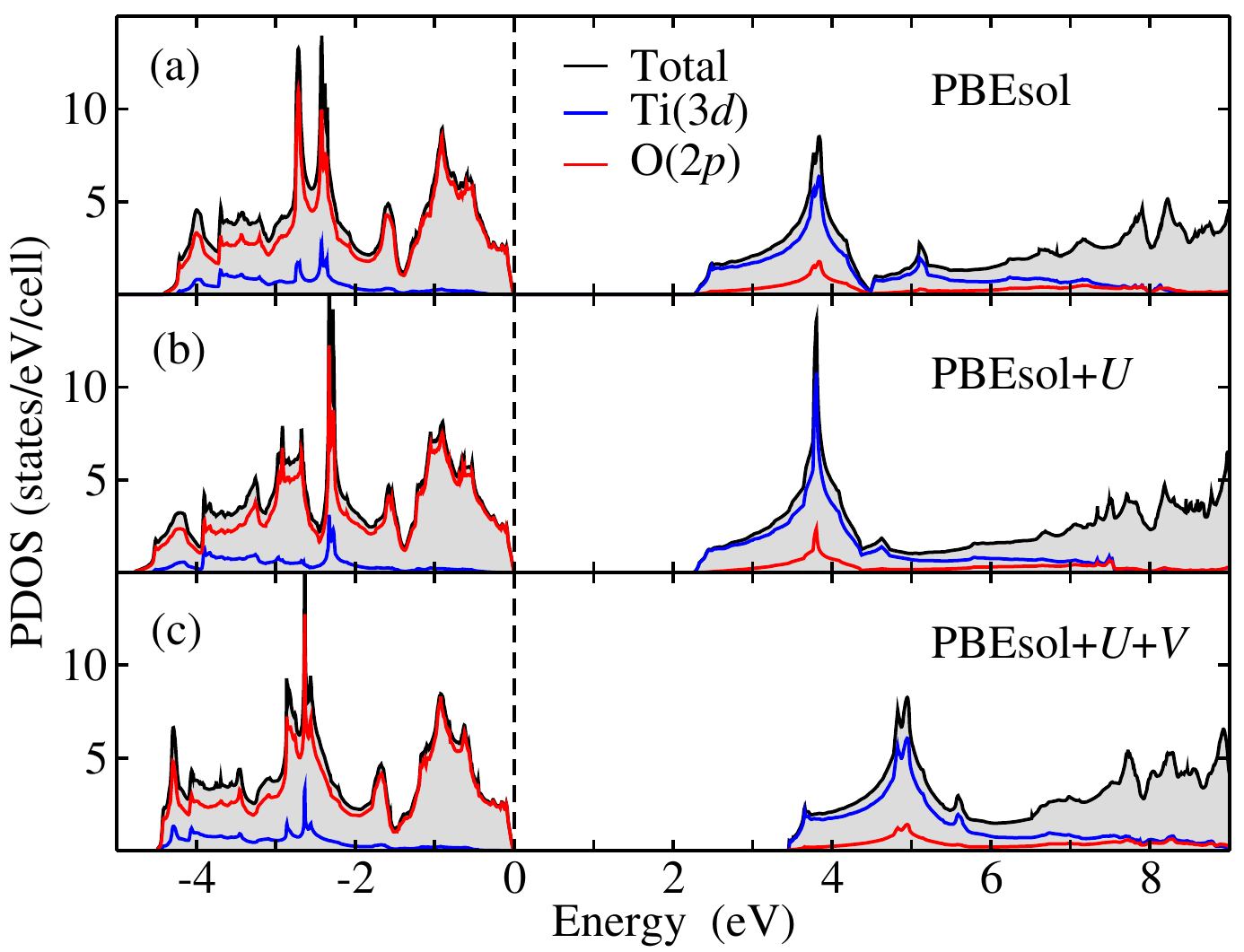}
	\caption{Computed total density of states and PDOS for Ti($3d$) and O($2p$) states using three functionals: (a)~PBEsol, (b)~PBEsol+$U$, and (c)~PBEsol+$U$+$V$. The zero of the energy is set at the top of the valence bands in all cases.}
	\label{fig:pdos_tot}
\end{figure}

Using the optimized structural parameters for each functional, here we analyze the respective PDOS and the band-gap values. Figure~\ref{fig:pdos_tot} illustrates the total DOS and PDOS for Ti($3d$) and O($2p$) states. It can be seen that the valence band maximum primarily originates from the O($2p$) states, while the conduction band minimum is mainly associated with the Ti($3d$) states. However, due to the hybridization between Ti($3d$) and O($2p$) states, there is some contribution from the Ti($3d$) states in the valence band region, and vice versa, the O($2p$) states contribute to the conduction band region. This hybridization effect is crucial for the covalency of the Ti-O bonds in BTO, as previously discussed in Sec.~\ref{sec:intro}. The nonvanishing contribution of Ti($3d$) states in the valence region supports our earlier assertion that applying the Hubbard $U$ correction to these states will not result in a negligible effect.  As can be seen in Figs.~\ref{fig:pdos_tot}(b) and \ref{fig:pdos_tot}(c), the application of Hubbard corrections introduces noticeable changes in the PDOS; the primary impact is on the band-gap value, and additionally there are some subtle adjustments in the shape of the PDOS. 

As discussed earlier, applying the Hubbard $U$ correction solely to the Ti($3d$) states leads to their increased localization and a reduction in their hybridization with the neighboring O($2p$) states. In the scenario where this hybridization is strongly suppressed, there would be vanishing energy overlap between Ti($3d$) and O($2p$) states. Consequently, the valence band region would solely consist of O($2p$) states, while the Ti($3d$) states would remain entirely unoccupied. Indeed, it can be noticed in Fig.~\ref{fig:pdos_tot}(b) that the application of $U$ does slightly reduce the Ti($3d$) weight in the valence region and increase it in the conduction band region, and the opposite effect is observed for the O($2p$) states. In fact, the integrated intensity of the Ti($3d$) states in the valence region is decreased by $\sim 10\%$ when applying our first-principles Hubbard $U$ correction. 

Let us now analyze the impact of the Hubbard corrections on the band gap. It is important to recall that DFT is not a theory for spectral properties. Importantly, in Ref.~\cite{KirchnerHall:2021} it is shown that DFT+$U$ with the Hubbard $U$ parameter determined using linear-response theory, can markedly enhance agreement with experimental band gaps. This improvement occurs when the Hubbard correction is applied to edge states (in cases where the system is already insulating at the DFT level) or states near the Fermi level (in situations where the system exhibits unphysical metallic behavior in standard DFT calculations), resulting in a Koopmans-like linearization~\cite{Nguyen:2018}.

Table~\ref{table:band-gab-comparison} summarizes our results and compares them with previous studies and experimental data. To the best of our knowledge, there is no direct experimental data available for the band gap of the rhombohedral BTO at low temperatures. However, the experimental band gap for the cubic phase is known to be 3.2~eV at 420~K, and it increases with decreasing temperature at a rate of 4.5 $\times$ 10$^{-4}$~eV/K~\cite{wemple}, leading to an extrapolated value of 3.4~eV at 0~K. We employ this extrapolated value for comparisons, noting that precise measurements of the rhombohedral BTO's band gap at low temperatures are required. We note that this extrapolated value was also used for comparisons in other works, e.g., in Refs.~\cite{Tsunoda_bandgap, Erhart:2007}. Our PBEsol calculation yields a band gap that underestimates the extrapolated experimental value by approximately 33\%. This is consistent with the trends observed in LDA~\cite{Huai-LDA} and PBE~\cite{Evarestov_LDA_PBE_PBE0} calculations. Hybrid functionals, such as HSE06~\cite{Tsunoda_bandgap} and PBE0~\cite{Evarestov_LDA_PBE_PBE0}, overestimate the band gap by approximately 9\% and 44\%, respectively. Thus, HSE06 has been the most accurate so far in predicting the band gap of rhombohedral BTO. Surprisingly, the band gap obtained with PBEsol+$U$ is very close to the PBEsol band gap. This result can be attributed to the cancellation of two effects: the decrease in the gap due to the vanishing rhombohedral distortion, and the increase in the gap caused by the application of the $U$ correction which generally leads to larger band gaps. To elucidate this, we performed a PBEsol calculation using the PBEsol+$U$ geometry (i.e., cubic structure), resulting in a band gap of 1.60~eV, while performing a PBEsol+$U$ calculation using the PBEsol+$U$ geometry yields a band gap of 2.26~eV. On the other hand, PBEsol+$U$+$V$ provides a band gap in excellent agreement with the extrapolated reference value, with a deviation of approximately 2\%. However, it is important to remark that the experimental band gap from Ref.~\cite{wemple} was determined through an optical absorption spectroscopy experiment, introducing the potential relevance of excitonic effects. Indeed, Ref.~\cite{Sanna-2011} demonstrated that incorporating excitonic effects (via solving the Bethe-Salpeter equation on top of $GW$) in the cubic and tetragonal phases of BTO induces a shift at the optical absorption onset, resulting in a band-gap reduction of approximately $0.3-0.5$ eV compared to $GW$ calculations. Consequently, a similar order of magnitude adjustment in the band gap might be expected for the rhombohedral phase of BTO due to excitonic effects. All the theoretical band gaps in Table~\ref{table:band-gab-comparison} exclude excitonic effects, implying an expected reduction on the order of $0.3-0.5$ eV. This adjustment aligns HSE06 more closely with the experimental band gap, whereas the PBEsol+$U$+$V$ band gap is likely to deviate more significantly from the experimental value. Overall, while PBEsol+$U$+$V$ does not exhibit significant improvements in structural properties compared to PBEsol (which already shows remarkable agreement with experiments), it significantly enhances the accuracy in predicting the band-gap value.

\begin{table}[t]
	\centering
    \renewcommand{\arraystretch}{1.5}
	\begin{tabular}{cccccccc}
        \hline\hline
        \multicolumn{3}{c}{This work}  & \parbox{0.95cm}{Ref.~$^a$} & \parbox{0.95cm}{Ref.~$^b$} & \parbox{0.95cm}{Ref.~$^c$} & \parbox{0.95cm}{Ref.~$^d$} & \parbox{0.95cm}{Ref.~$^e$}  \\
         PBEsol  &  +$U$  & +$U$+$V$   &   LDA     &   PBE     &   HSE06   &  PBE0     & Expt.      \\ \hline
          2.28   &  2.26  &   3.46     &   2.23    &   2.7     &   3.69    &  4.9      & 3.4        \\
        \hline\hline
	\end{tabular}
	\caption{Comparison of the band gap (in eV) computed in this work using PBEsol, PBEsol+$U$, and PBEsol+$U$+$V$ and in previous works using (semi-)local and hybrid functionals and as measured in experiments. The experimental band gap was estimated using the extrapolation technique as explained in the main text. Ref.~$^a$ is \cite{Huai-LDA}, Ref.~$^b$ is \cite{Evarestov_LDA_PBE_PBE0}, Ref.~$^c$ is \cite{Tsunoda_bandgap}, Ref.~$^d$ is \cite{Evarestov_LDA_PBE_PBE0}, Ref.~$^e$ is \cite{wemple}.}
	\label{table:band-gab-comparison}
\end{table}

\subsection{Born effective charges and dielectric tensor}
\label{BEC-Diel}

In this section, we present and discuss the computed BEC and dielectric tensor using the three functionals considered in this work. Table~\ref{table:BEC} provides a comparison of our calculations with previous computational studies~\cite{Ghosez_BEC, Hermet_2009, Seo-BEC-Diel-B1WC}. In the rhombohedral crystal structure, the computed $Z^{*}_{\mathrm{Ti}}$ and $Z^{*}_{\mathrm{Ba}}$ are diagonal tensors with $xx$ and $yy$ components being equal due to symmetry, while the BEC of three O atoms have nonzero off-diagonal components. To simplify the analysis, we diagonalize $Z^{*}_{\mathrm{O}}$ and denote its largest and doubly-degenerate smallest eigenvalues as O$_{||}$ and O$_{\perp}$, respectively (which are the same for three O atoms). O$_{||}$ represents displacements of O ions along the Ti$-$O bond, while O$_{\perp}$ refers to displacements perpendicular to the bond. As shown in Table~\ref{table:BEC}, for all functionals, the values of $Z^{*}_{\mathrm{Ti}}$ diagonal components and $Z^{*}_{\mathrm{O}_{||}}$ are substantially larger than their nominal values ($+4$ and $-2$, respectively)~\cite{Ghosez_BEC, Mellaers-PBEsol-22}. This observation is consistent with other ABO$_3$ perovskites and is a fingerprint of covalency of the Ti$–$O bonds~\cite{Cohen_nature, Ghosez_BEC, GOH2016-BEC, Axe-67, Seo-BEC-Diel-B1WC}. In contrast, the computed $Z^{*}_{\mathrm{Ba}}$ and $Z^{*}_{\mathrm{O}_\perp}$ values are not significantly different from their nominal values ($+2$ for Ba), suggesting an ionic character of the Ba$–$O bonds~\cite{Ghosez_BEC, Seo-BEC-Diel-B1WC, GOH2016-BEC}. Applying the onsite $U$ correction alone does not lead to large changes in the BEC for all the elements as compared to PBEsol (except for $Z^{* zz}_{\mathrm{Ti}})$. When considering PBEsol+$U$+$V$, however, we find that $Z^{* xx}_{\mathrm{Ti}}$, $Z^{* zz}_{\mathrm{Ti}}$, and $Z^{*}_{\mathrm{O}_{||}}$ are smaller by $7-13\%$ compared to PBEsol. This reduction in the BEC due to $+U$+$V$ corrections arises from the redistribution of electronic charge between Ti and O ions and sensitivity to changes in structural parameters (see Table~\ref{table:structural-optimization})~\cite{Ghosez_BEC}.

\begin{table}[t]
    \renewcommand{\arraystretch}{1.2}
	\begin{center}
		\begin{tabular}{lcccccc}
			\hline\hline
Functional                    & \parbox{0.9cm}{$Z^{* xx}_{\mathrm{Ba}}$}  & \parbox{0.9cm}{$Z^{* zz}_{\mathrm{Ba}}$} & 
                                \parbox{0.9cm}{$Z^{* xx}_{\mathrm{Ti}}$}  & \parbox{0.9cm}{$Z^{* zz}_{\mathrm{Ti}}$} & 
                                \parbox{0.9cm}{$Z^{*}_{\mathrm{O}_{||}}$} & \parbox{0.9cm}{$Z^{* }_{\mathrm{O}_\perp}$}   \\ \hline	

PBEsol                        & 2.79      & 2.75      & 6.61      & 5.72  &  $-5.11$  & $-1.99$ \\
PBEsol+$U$                    & 2.77      & 2.77      & 6.57      & 6.56  &  $-5.20$  & $-2.07$ \\
PBEsol+$U$+$V$                & 2.80      & 2.74      & 6.18      & 5.08  &  $-4.47$  & $-1.95$ \\
LDA~\cite{Ghosez_BEC}         & 2.79      & 2.74      & 6.54      & 5.61  &  $-5.05$  & $-1.98$ \\
LDA~\cite{Hermet_2009}        & 2.78      & 2.74      & 6.61      & 5.77  &  $-5.10$  & $-1.99$ \\			
B1-WC~\cite{Seo-BEC-Diel-B1WC}& 2.71      & 2.68      & 6.41      & 5.55  &  $-4.94$  & $-1.95$ \\
			\hline\hline
		\end{tabular}
	\end{center}
	\caption{Comparison of the computed BEC of Ba, Ti, and O using PBEsol, PBEsol+$U$, and PBEsol+$U$+$V$ and those from previous studies~\cite{Ghosez_BEC, Hermet_2009, Seo-BEC-Diel-B1WC}. The $xx$ and $yy$ components are equal for $Z^{*}_{\mathrm{Ba}}$  and $Z^{*}_{\mathrm{Ti}}$, hence  we report only the former. The BEC values in Refs.~\cite{Ghosez_BEC} and \cite{Hermet_2009} are calculated at experimental lattice parameters~\cite{kwei}, with relaxed atomic positions.}
	\label{table:BEC}
\end{table} 

Let us now discuss the dielectric constant of the low-temperature rhombohedral BTO. Again due to symmetry, the $xx$ and $yy$ components of the dielectric tensor are equal. In Table~\ref{table:dieletric-tensor}, we present a comparison of our computed values with those from previous studies~\cite{Evarestov_LDA_PBE_PBE0, Mellaers-PBEsol-22, Ghosez, Paoletta_Diele-21, Seo-BEC-Diel-B1WC} and experiments~\cite{Eltes-Diel-Exp}. First, we want to remark that quite often~\cite{Wu-Xifan_LDA, Evarestov_LDA_PBE_PBE0, Seo-BEC-Diel-B1WC, Seo2013} in the literature the theoretical components of the dielectric tensor of the rhombohedral BTO are compared with the following experimental values: $\epsilon^{\infty}_{xx} = 6.19$ and $\epsilon^{\infty}_{zz} = 5.88$~\cite{Wang:2001}. However, as was rightly pointed out in Ref.~\cite{Mahmoud-PBE0}, the experimental data of Ref.~\cite{Wang:2001} corresponds to the room-temperature tetragonal phase of BTO, and thus it should not be used for comparisons with the low-temperature rhombohedral phase. On the other hand, we follow Ref.~\cite{Paoletta_Diele-21} and use an average experimental dielectric constant $\langle \epsilon \rangle = 5.20$ which was obtained from the average refractive index of 2.28~\cite{Eltes-Diel-Exp}. However, it is crucial to emphasize that more accurate measurements of the dielectric tensor components for the low-temperature rhombohedral phase of BTO are needed. As can be seen in Table~\ref{table:dieletric-tensor}, our PBEsol dielectric tensor shows good agreement with previous LDA~\cite{Ghosez_BEC} and PBEsol~\cite{Mellaers-PBEsol-22} studies. The average dielectric constant obtained using PBEsol overestimates the experimental value by $\sim 15\%$. When including the $+U$ correction, $\langle \epsilon \rangle$ slightly decreases, but it still overestimates the experimental value by $\sim 13\%$. However, it is essential to consider that the PBEsol+$U$ geometry corresponds to the cubic phase at 0~K, making this comparison less straightforward. We note in passing that the experimental $\langle \epsilon \rangle$ of the cubic phase of BTO is 5.40~\cite{BURNS19829}. The PBEsol+$U$+$V$ functional yields $\langle \epsilon \rangle$ that is in substantially closer agreement with the experimental value, overestimating it by only $\sim 3\%$. Hence, PBEsol+$U$+$V$ provides the most accurate prediction for $\langle \epsilon \rangle$ compared to PBEsol and PBEsol+$U$. Finally, we note that hybrid functionals PBE0~\cite{Evarestov_LDA_PBE_PBE0} and B1-WC~\cite{Seo-BEC-Diel-B1WC} underestimate $\langle \epsilon \rangle$ by $\sim 14\%$ and $\sim 7\%$, respectively, while $\langle \epsilon \rangle$ from PBE~\cite{Evarestov_LDA_PBE_PBE0} is by chance in excellent agreement with the experimental value.

\begin{table}[t]
 	\centering
    \renewcommand{\arraystretch}{1.2}
 	\begin{tabular}{lccc}
        \hline\hline
 		Functional & \parbox{1.87cm}{$\epsilon^{\infty}_{xx}$} & \parbox{1.87cm}{$\epsilon^{\infty}_{zz}$} & \parbox{1.87cm}{$\langle \epsilon \rangle$} \\ \hline
 		PBEsol                                             & 6.14 &  5.69 & 5.99 \\ 
        PBEsol+$U$                                         & 5.88 &  5.88 & 5.88 \\
 	 	PBEsol+$U$+$V$                                     & 5.54 &  5.00 & 5.36 \\ 
 		LDA \cite{Ghosez}                                  & 6.16 &  5.69 & 6.00 \\ 
        LDA \cite{Evarestov_LDA_PBE_PBE0}                  & 6.05 &  5.82 & 5.97 \\
 		LDA+$\Delta$ \cite{Paoletta_Diele-21} & 5.57 &  5.51 & 5.55 \\
        PBE  \cite{Evarestov_LDA_PBE_PBE0}                 & 5.42 &  4.80 & 5.21 \\
        PBEsol~\cite{Mellaers-PBEsol-22}                   & 6.11 &  5.63 & 5.95 \\
        PBE0 \cite{Evarestov_LDA_PBE_PBE0}                 & 4.62 &  4.13 & 4.46 \\
 		B1-WC \cite{Seo-BEC-Diel-B1WC}                     & 4.96 &  4.60 & 4.84 \\
        Expt.~\cite{Eltes-Diel-Exp}                        &      &       & 5.20 \\
        \hline\hline
 	\end{tabular}
 	\caption{Diagonal components of the dielectric tensor $\epsilon^{\infty}_{xx}$ and $\epsilon^{\infty}_{zz}$ (note that $\epsilon^{\infty}_{yy} = \epsilon^{\infty}_{xx}$) and its average value ($\langle \epsilon \rangle = (\epsilon^{\infty}_{xx} + \epsilon^{\infty}_{yy} + \epsilon^{\infty}_{zz})/3$) as computed in this work using PBEsol, PBEsol+$U$, PBEsol+$U$+$V$ and as obtained in previous works~\cite{Evarestov_LDA_PBE_PBE0, Mellaers-PBEsol-22, Ghosez, Paoletta_Diele-21, Seo-BEC-Diel-B1WC} and measured in experiments~\cite{Eltes-Diel-Exp}. In Ref.~\cite{Paoletta_Diele-21} a scissor $\Delta$ correction was used.}
 	\label{table:dieletric-tensor}
 \end{table}

\subsection{Phonons}
\label{phonon-dispersion} 

\begin{table*}[t]
 	\renewcommand{\arraystretch}{1.2}
    \centering
    \begin{tabular}{lllcccccccccc} 
	\hline\hline
\multirow{2}{*}{Mode} & \multirow{2}{*}{Label} & \multirow{2}{*}{\,\,\,$\mathbf{q}$} & \parbox{1.5cm}{PBEsol} & PBEsol+$U$+$V$ & PBEsol+$U$+$V$ & \parbox{1.4cm}{Ref.~$^a$} & \parbox{1.4cm}{Ref.~$^b$} & \parbox{1.4cm}{Ref.~$^a$} & \parbox{1.4cm}{Ref.~$^a$} & \parbox{1.4cm}{Ref.~$^c$} \\
               &      &          &  @PBEsol  & @PBEsol    & @PBEsol+$U$+$V$& LDA       &  LDA      & PBE       & PBE0      & B1-WC     \\ \hline
    TO         &      &          &           &            &                &           &           &           &           &           \\
    1$E$       & TO1  & $x$, $z$ &  166      &  168       & 162            & 145       &  163      & 165       &  180      &  125      \\ 
	1$A_{1}$   & TO1  & $x$      &  170      &  171       & 165            & 191       &  167      & 167       &  181      &  192      \\ 
	2$E$       & TO2  & $x$, $z$ &  207      &  243       & 264            & 191       &  210      & 264       &  264      &  217      \\  
	3$E$       & TO3  & $x$, $z$ &  294      &  290       & 288            & 306       &  293      & 299       &  319      &  318      \\
	2$A_{1}$   & TO2  & $x$      &  262      &  292       & 310            & 200       &  259      & 309       &  332      &  285      \\
	4$E$       & TO4  & $x$, $z$ &  472      &  483       & 478            & 489       &  470      & 474       &  493      &  497      \\
	3$A_{1}$   & TO3  & $x$      &  516      &  537       & 548            & 508       &  512      & 544       &  571      &  542      \\ \hline
    LO         &      &          &           &            &                &           &           &           &           &           \\
    1$E$       & LO1  & $x$      &  176      &  180       & 177            & 190       &  174      & 176       &  190      &  193      \\ 
	1$A_{1}$   & LO1  & $z$      &  180      &  186       & 183            & 193       &  178      & 182       &  197      &  199      \\ 
	2$E$       & LO2  & $x$, $z$ &  294      &  290       & 288            & 306       &  293      & 299       &  319      &  318      \\ 
	3$E$       & LO3  & $x$      &  441      &  443       & 437            & 460       &  441      & 441       &  468      &  472      \\
	2$A_{1}$   & LO2  & $z$      &  460      &  464       & 463            & 471       &  461      & 467       &  498      &  491      \\
    3$A_{1}$   & LO3  & $z$      &  681      &  705       & 701            & 707       &  676      & 689       &  739      &  740      \\
    4$E$       & LO4  & $x$      &  691      &  714       & 711            & 713       &  687      & 705       &  750      &  730      \\ \hline
    $A_{2}$    &      & $x$, $z$ &  277      &  275       & 270            & 297       &  277      & 279       &  301      &  307      \\ 
	\hline\hline
    \end{tabular}
	\caption{Phonon frequencies of the rhombohedral BTO (in cm$^{-1}$) at the Brillouin zone center computed using PBEsol on top of the PBEsol geometry (PBEsol@PBEsol), and PBEsol+$U$+$V$ on top of the PBEsol geometry (PBEsol+$U$+$V$@PBEsol) and on top of the PBEsol+$U$+$V$ geometry (PBEsol+$U$+$V$@PBEsol+$U$+$V$). The results from previous computational studies (Ref.~$^a$ is \cite{Evarestov_LDA_PBE_PBE0}, Ref.~$^b$ is \cite{Hermet_2009}, and Ref.~$^c$ is \cite{Seo2013}) are also shown. The symmetry of the phonon modes is specified in the first column, while the common labeling~\cite{Hermet_2009} of the TO and LO modes is specified in the second column. The label ``$\mathbf{q}$'' denotes the direction of the phonon wavevector $\mathbf{q}$, which is either along the $x$ or $z$ axis in reciprocal space of the rhombohedral BTO reference system (see Sec.~\ref{sec:comput_details}). The $3E$(TO3) and $2E$(LO2) modes are degenerate, while the $A_2$ mode is silent.}
	\label{table:Gamma-point-freq}
\end{table*}

Here we explore the lattice vibrational properties of BTO using the three considered functionals. A group-theory analysis reveals that the zone-center phonon frequencies of rhombohedral BTO can be decomposed as $\Gamma_\mathrm{opt} = 3 A_{1} + A_{2} + 4E$~\cite{Wu-Xifan_LDA}, where both $A_{1}$ and $E$ modes are infrared and Raman active, while the $A_{2}$ mode is silent. The zone-center phonon frequencies obtained from PBEsol and PBEsol+$U$+$V$ are compared with previous computational studies~\cite{Evarestov_LDA_PBE_PBE0, Hermet_2009, Seo2013} in Table~\ref{table:Gamma-point-freq}. 

In order to scrutinize the impact of lattice geometry ($a$ and $\alpha$) on phonon frequencies, we present the outcomes of PBEsol+$U$+$V$ phonon calculations for both the PBEsol+$U$+$V$ optimized geometry (referred to as PBEsol+$U$+$V$@PBEsol+$U$+$V$) and the PBEsol optimized geometry (PBEsol+$U$+$V$@PBEsol). It is worth noting that in both cases, atomic positions are optimized using PBEsol+$U$+$V$ to ensure that forces acting on atoms are vanishing. Conversely, PBEsol+$U$ is omitted from this comparison due to its inability to sustain the rhombohedral phase of BTO (refer to Sec.~\ref{structural}). The nonanalytic component of the dynamical matrix up to the dipolar order is taken into account, which leads to the splitting between the longitudinal optical (LO) and transverse optical (TO) modes~\cite{baroni_phonons_2001}. This nonanalytic component is derived using the BEC and dielectric tensor, as discussed in Sec.~\ref{BEC-Diel}. In the case of PBEsol+$U$+$V$@PBEsol, the BEC and dielectric tensor are not recalculated; instead, those from the PBEsol+$U$+$V$@PBEsol+$U$+$V$ case are used.  

The phonon mode data presented in Table~\ref{table:Gamma-point-freq} demonstrates distinctive behaviors among the considered functionals in this study and in previous works. We have not compared our zone-center phonons to the experimental data taken on single crystals in Refs.~\cite{Tenne-Ramman, Maslova2020} as these samples are polydomain with the $z$ axis oriented along the $c$ axis of the tetragonal BTO reference system. This reorientation leads both to a mixing of the phonon modes listed in Table~\ref{table:Gamma-point-freq} (see Secs.~\ref{sec:comput_details} and \ref{Raman} for more details) and to potential differences in the resulting LO-TO splitting. Experimental data for rhombohedral BTO single crystals with a single ferroelectric domain, specifically for $\mathbf{q}$ directions along the $x$ and $z$ axes in reciprocal space of the rhombohedral BTO reference system (where, in real space, $z$ is parallel to the $C_3$ rhombohedral axis), would be highly desirable for evaluating the accuracy of computational data obtained from various functionals, as presented in Table~\ref{table:Gamma-point-freq}.

\begin{figure*}[t]
	\centering
	\includegraphics[width=0.47\textwidth]{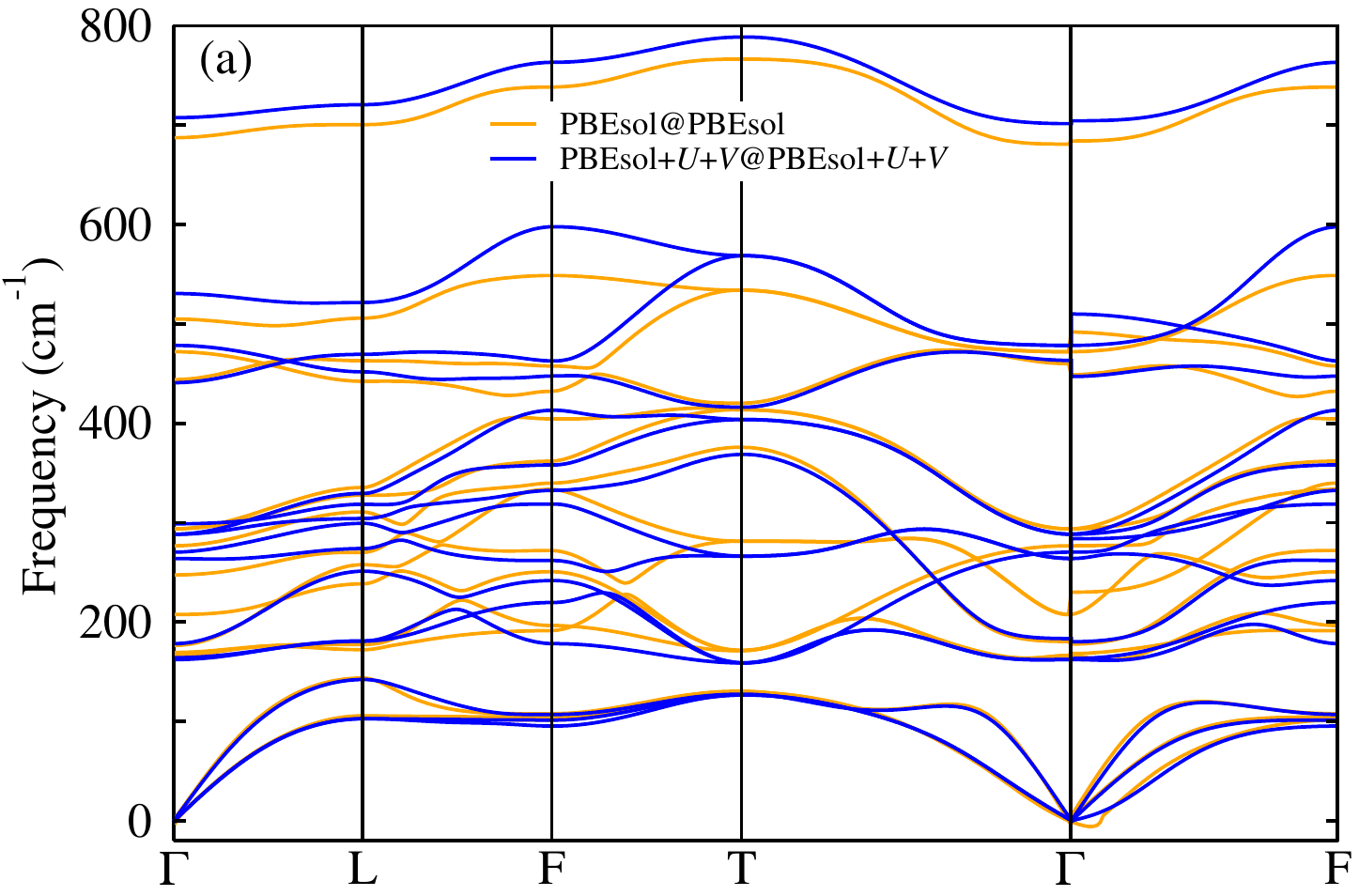}
    \hspace{0.2cm}
	\includegraphics[width=0.47\textwidth]{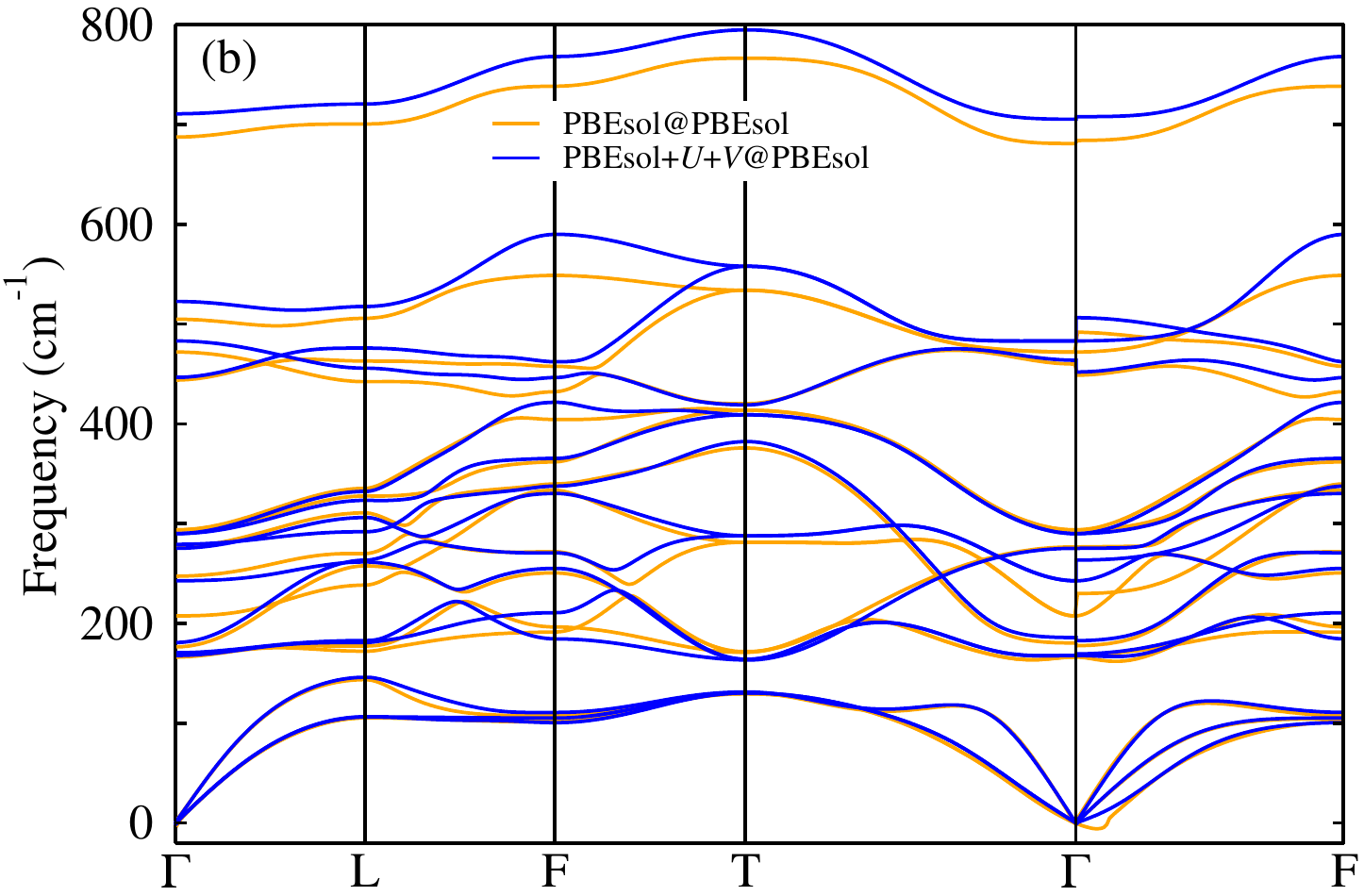}
	\caption{(a)~Phonon dispersion of the rhombohedral BTO computed using PBEsol and PBEsol+$U$+$V$, and using different optimized geometries. (a)~PBEsol+$U$+$V$ calculation using the PBEsol+$U$+$V$ lattice geometry (PBEsol+$U$+$V$@PBEsol+$U$+$V$), and (b)~PBEsol+$U$+$V$ calculation using the PBEsol lattice geometry (PBEsol+$U$+$V$@PBEsol). On both panels the PBEsol results are exactly the same and were obtained using the PBEsol lattice geometry.}
	\label{fig:phonon_dispersion}
\end{figure*}

Let us now shift to the phonon dispersion of the rhombohedral BTO lattice. In Fig.~\ref{fig:phonon_dispersion}, we present a comprehensive comparison of the computed phonon frequencies as a function of the phonon wave vector $\mathbf{q}$ along high-symmetry directions in the Brillouin zone obtained from PBEsol@PBEsol, PBEsol+$U$+$V$@PBEsol+$U$+$V$, and PBEsol+$U$+$V$@PBEsol. It can be seen that there is a discontinuity in the phonon dispersion near $\Gamma$ for certain optical phonon modes, due to the LO-TO splitting that depends on the direction of the phonon wave vector $\mathbf{q}$ in the long-wavelength limit ($\mathbf{q} \rightarrow \mathbf{0}$)~\cite{baroni_phonons_2001}. The three acoustic branches display remarkable congruence across all cases, except in proximity to the $\Gamma$ point. This correspondence stems from the fact that, up to around $150$cm$^{-1}$, the primary contribution to the vibrational density of states (vDOS) can be attributed to the motion of Ba ions~\cite{Huai-LDA}, which remains largely unaffected by the Hubbard corrections. However, above $\sim 150$~cm$^{-1}$, the vDOS is mainly influenced by both Ti and O atoms, leading to pronounced differences in the phonon bands due to the Hubbard corrections. Specifically, within PBEsol+$U$+$V$@PBEsol+$U$+$V$ and PBEsol+$U$+$V$@PBEsol, the highest phonon bands are shifted to higher frequencies (compared to PBEsol@PBEsol), consistently with similar observations in other materials~\cite{Floris:2011, Floris:2020}. The effect of the lattice parameters on the PBEsol+$U$+$V$ phonon calculations can be seen by comparing Figs.~\ref{fig:phonon_dispersion}(a) and \ref{fig:phonon_dispersion}(b). Although PBEsol+$U$+$V$@PBEsol+$U$+$V$ and PBEsol+$U$+$V$@PBEsol phonon dispersions exhibit qualitative similarity, notable quantitative disparities exist. It is important to pay special attention to the behavior of the acoustic phonon bands near the $\Gamma$ point. As previously noted in Ref.~\cite{Massimiliano}, piezoelectric materials like BTO exhibit imaginary phonon frequencies around $\Gamma$ when considering the nonanalytic part of the dynamical matrix only up to the dipolar order. To address this issue, higher multipolar orders of the nonanalytic correction to the dynamical matrix must be included~\cite{Massimiliano}, which are though not considered in our study. In our calculations, we find sizable imaginary phonon frequencies along the $\Gamma$-F high-symmetry direction when using the PBEsol functional. However, these imaginary frequencies are substantially reduced when using PBEsol+$U$+$V$, becoming imperceptible on the plot due to the large frequency scale. Such differences between the PBEsol and PBEsol+$U$+$V$ results do not arise from the variations in the lattice geometry, because both PBEsol+$U$+$V$@PBEsol+$U$+$V$ and PBEsol+$U$+$V$@PBEsol do not have a large negative bump along $\Gamma$-F like in the PBEsol@PBEsol case. Hence, such differences between PBEsol and PBEsol+$U$+$V$ originate purely from changes in the electronic structure due to the $U$ and $V$ corrections. Furthermore, both PBEsol and PBEsol+$U$+$V$ yield minimal imaginary phonon frequencies close to $\Gamma$ along the $\Gamma$-L direction, although these frequencies are not readily visible on the large frequency scale shown in Fig.~\ref{fig:phonon_dispersion}. These observations suggest that higher-order corrections to the dynamical matrix are essential in BTO and similar materials, regardless of the functional employed.

As was shown in Ref.~\cite{Zhong:1994} for the cubic BTO, there is a giant LO-TO splitting. Conversely, in the rhombohedral BTO we find that there is a strong mixing of LO modes when incorporating the nonanalytic part to the dynamical matrix. Therefore, there is no one-to-one connection between the highest LO mode and one of the TO modes, and hence it is not straightforward how to determine the LO--TO splitting. Nevertheless, we believe that it is useful to point out what is the optical phonon bandwidth, i.e., the difference between the highest LO and lowest TO modes, which is 525~cm$^{-1}$ in PBEsol and about 550~cm$^{-1}$ in PBEsol+$U$+$V$.

\begin{figure}[t]
	\centering
	\includegraphics[width=0.47\textwidth]{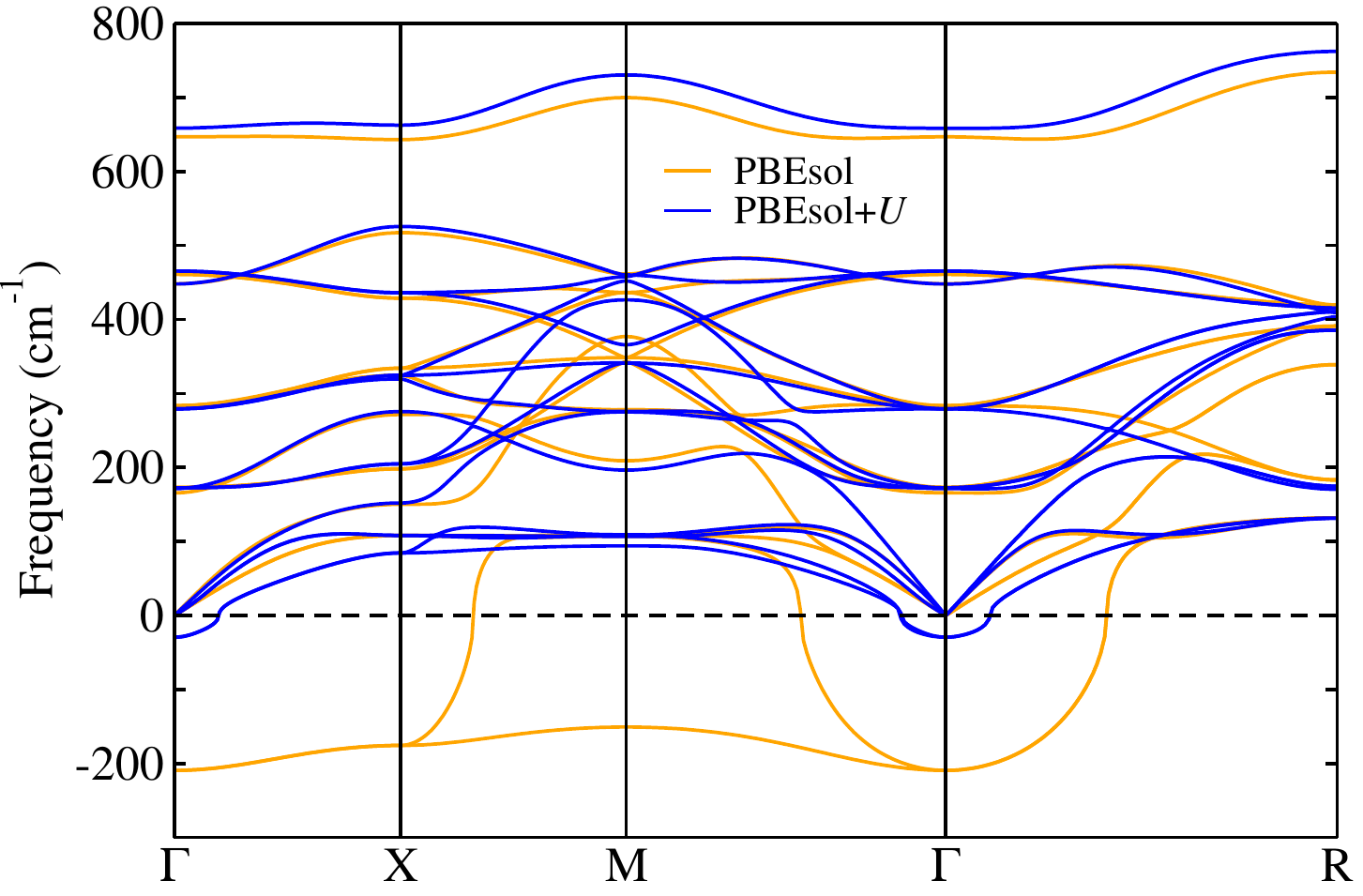}
	\caption{Phonon dispersion of the cubic BTO computed using PBEsol and PBEsol+$U$, both using the PBEsol+$U$ lattice geometry.}
	\label{fig:phonon_dispersion_cubic}
\end{figure}

It is instructive to compare the phonon dispersion for the cubic phase of BTO using PBEsol and PBEsol+$U$. Figure~\ref{fig:phonon_dispersion_cubic} shows such a comparison when using the PBEsol+$U$ lattice geometry for both functionals (i.e., PBEsol@PBEsol+$U$ and PBEsol+$U$@PBEsol+$U$, respectively). In our PBEsol calculations for the cubic BTO, we observe large imaginary phonon frequencies at various points in the Brillouin zone, including the high-symmetry points $\Gamma$, M, and X, which is consistent with previous studies~\cite{Zhang-BECs-Diele-2017, Xie_2008, Michele-BOT-22, Ghosez:1998}. When employing PBEsol+$U$ with the first-principles $U$ value, the imaginary phonon frequencies vanish everywhere in the Brillouin zone except in the vicinity of the $\Gamma$ point. Since soft phonon modes measure the structural instability, this result shows that PBEsol+$U$ tends to stabilize the cubic phase at 0~K. A similar observation was reported in Ref.~\cite{Tsunoda_bandgap} (see Fig.~S6 in this reference). However, it is worth noting that no imaginary phonon frequencies were found in that study, even around $\Gamma$. Despite the fact that the value of the Hubbard $U$ parameter is extremely similar in both studies (4.52~eV in our case vs 4.49~eV in Ref.~\cite{Tsunoda_bandgap}), the remaining differences might be attributed to the following factors: $i)$~we use PBEsol+$U$ in this study, while PBE+$U$ was used in Ref.~\cite{Tsunoda_bandgap}, $ii)$~we employ L\"owdin-orthogonalized atomic orbitals, while in Ref.~\cite{Tsunoda_bandgap}, the PAW functions were used to build the Hubbard projectors, $iii)$~different pseudopotentials were used in both studies, and $iv)$~we use the PBEsol+$U$ optimized lattice parameter of 3.990~\AA\ (see Table~\ref{table:structural-optimization}), whereas in Ref.~\cite{Tsunoda_bandgap}, a lattice parameter of 4.077~\AA\ was used. Although it is not clear which factor plays the dominant role, the conclusion remains that soft phonon modes around $\Gamma$ are highly sensitive to numerical details of calculations. Additionally, we verified that using larger values of the $U$ parameter (e.g., testing with $U=6$~eV) within our computational setup completely eliminates the soft phonon modes around $\Gamma$.

Finally, we address the comparison between the tendency of PBEsol+$U$ to stabilize the cubic phase of BTO at 0~K and the effect of strain within PBEsol to achieve a similar stabilization. Previous studies have shown that by compressing the lattice (reducing the lattice parameter by $\sim 3\%$), the cubic phase of BTO can be stabilized at 0~K when using (semi-)local functionals since no imaginary phonon frequencies are present~\cite{Michele-BOT-22, Xie_2008, Seo-BEC-Diel-B1WC}. This phenomenon can be explained as follows: When BTO is compressed, the interatomic distances decrease, altering the energy landscape. Specifically, the minima of the symmetric double well of the energy potential move closer together, while the saddle point becomes lower, resulting in the stabilization of the high-symmetry cubic phase. On the other hand, when applying the first-principles $U$ correction, the lattice tends to expand, as expected (see the discussion in Sec.~\ref{structural}). We found that the optimized lattice parameter for the cubic BTO using PBEsol+$U$ is increased only by $\sim 0.3\%$ compared to PBEsol~\footnote{The optimized lattice parameter for the cubic BTO at 0~K using PBEsol is found to be 3.979~\AA.}. Therefore, from a purely geometrical perspective, it is not evident why PBEsol+$U$ would stabilize the cubic phase of BTO. As we discussed earlier, the reason lies purely in the electronic behavior. The localization of the Ti($3d$) states suppresses the Ti($3d$)--O($2p$) hybridization that is crucial for the covalency of the Ti-O bonds and stabilization of the rhombohedral phase of BTO. Consequently, as the strength of the Hubbard $U$ correction is increased, the dynamical stability of the rhombohedral phase diminishes, while that of the cubic phase increases. Thus, this intricate electronic effect is responsible for the observed stabilization of the cubic phase by PBEsol+$U$.

\subsection{Raman spectra}
\label{Raman}

In this section, we present a comparison between the computed and experimental nonresonant Raman spectra of rhombohedral BTO. The spectra were experimentally measured at 10~K~\cite{Tenne-Ramman} and 80~K~\cite{Buscaglia, Maslova2020}. Although all these experimental spectra display similar qualitative features, we choose to compare with the higher-resolution Raman spectrum on single crystals of BTO from Ref.~\cite{Maslova2020}. It is important to note that the Raman spectrum of single crystals in Ref.~\cite{Maslova2020} closely resembles the spectrum presented in Ref.~\cite{Tenne-Ramman}. The latter reference explicitly states that the measurements were conducted using the backscattering geometry with the $z(xx)\bar{z}$ setup, where $z$ is along the $c$ axis of the tetragonal BTO phase. We use the same setup in our calculations, as detailed in Sec.~\ref{sec:comput_details}. On the other hand, we are aware of only two theoretical studies of the Raman spectra of the rhombohedral BTO, and they are based on DFPT using the LDA functional~\cite{Hermet_2009, Popov:2020}. In Ref.~\cite{Hermet_2009} the authors considered the $x(zz)y$ and $z(xy)\bar{z}$ setups for single crystals with the $z$ axis aligned along the $C_3$ rhombohedral axis, while in Ref.~\cite{Popov:2020} the averaged Raman spectra of polycrystalline rhombohedral BTO were computed. Hence, the computational predictions of these studies are not directly comparable with our Raman spectrum calculations for single crystals and the experiments of Refs.~\cite{Tenne-Ramman, Maslova2020}. 

\begin{figure}[t]
  \includegraphics[width=0.45\textwidth]{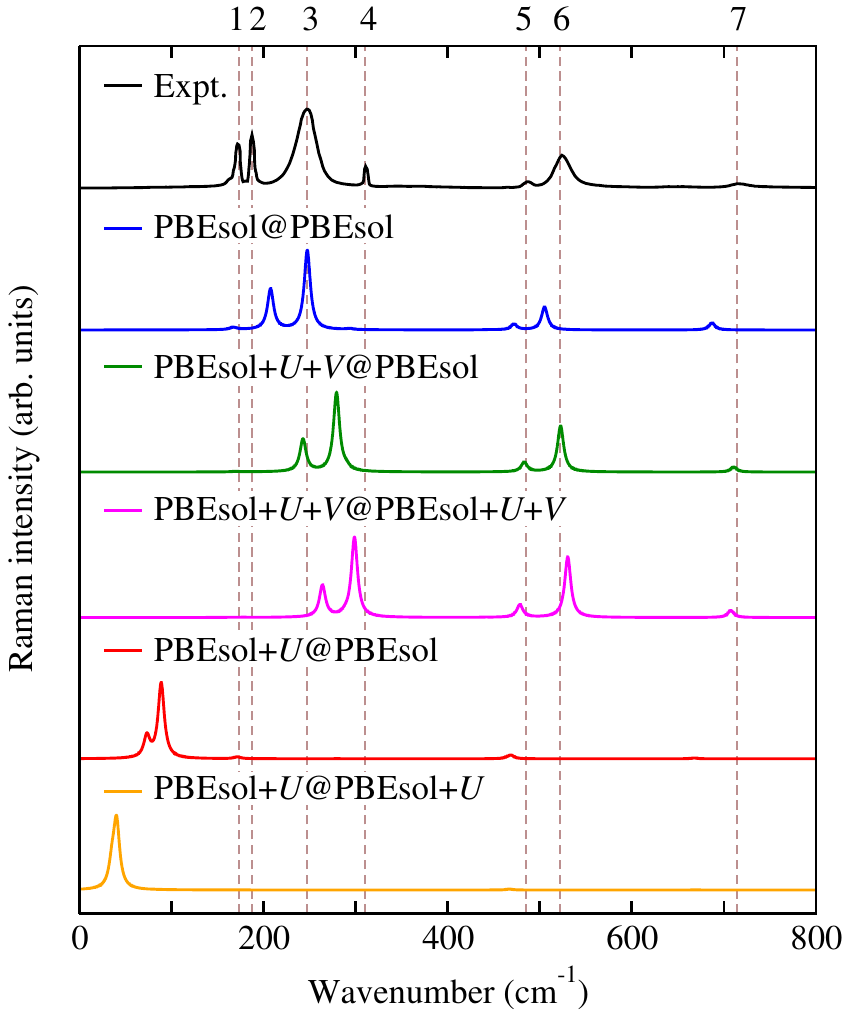}
  \caption{Raman spectra computed using PBEsol, PBEsol+$U$, and PBEsol+$U$+$V$ on top of different optimized geometries. The experimental Raman spectra were obtained for single crystals at 80~K~\cite{Maslova2020}. All spectra are shifted vertically for the sake of clarity. Vertical dashed lines correspond to positions of peaks in the experimental Raman spectrum, and they are numbered for convenience.}
\label{fig:raman-spectra}  
\end{figure}

Figure~\ref{fig:raman-spectra} illustrates the computed Raman spectra using the three considered functionals along with the experimental spectrum from Ref.~\cite{Maslova2020}. The Hubbard corrections exert a twofold influence on the Raman spectra: $i)$~via alterations in the optimized lattice geometry and $ii)$~via changes in the electronic structure. To disentangle these distinct effects, separate Raman spectrum computations were executed for PBEsol+$U$ and PBEsol+$U$+$V$, each employing two distinct lattice geometries: one optimized utilizing PBEsol+$U$ and PBEsol+$U$+$V$, correspondingly, and the other using the PBEsol optimized lattice geometry.

We begin our analysis of Fig.~\ref{fig:raman-spectra} by comparing the Raman spectrum obtained from PBEsol@PBEsol with the experimental one. The peak~3 is well reproduced (its position coincides very well with the experimental one), peak~2 is red-shifted by 19~cm$^{-1}$, peak~1 is barely visible and it is blue-shifted by 5 cm$^{-1}$, while peak~4 is not visible at all. Next, peaks~5, 6, and 7 are all blue-shifted by 15, 19, and 28 cm$^{-1}$, respectively. We recall that here we use a constant broadening with the Lorenzian function, hence we do not account for the various lifetimes of all peaks, which might explain the differences in the widths of the peaks. The differences in the positions of peaks between PBEsol@PBEsol and experiments might be attributed to the following factors: $i)$~errors stemming from the approximations when using PBEsol@PBEsol, and $ii)$~our calculations are performed assuming a single ferroelectric domain while measurements in Ref.~\cite{Maslova2020} were carried out on a polydomain single crystal sample. Overall, our PBEsol@PBEsol spectrum looks qualitatively similar to the experimental spectrum, although some disparities in peak positions and relative intensities persist. 

Let us now delve into the analysis of the Raman spectra computed employing the Hubbard corrections. As depicted in Fig.~\ref{fig:raman-spectra}, the Raman spectrum computed using PBEsol+$U$+$V$ with the PBEsol+$U$+$V$ optimized lattice geometry (PBEsol+$U$+$V$@PBEsol+$U$+$V$) displays a notable deterioration in comparison to PBEsol@PBEsol and the experiment. Namely, peaks~2 and 3 are red-shifted even more, peak~5 does not show significant changes, peak~6 is now red-shifted with respect to the experimental peak, while peak~7 is now in much closer agreement with the experimental peak position. A substantial improvement emerges when the Raman spectrum is calculated using PBEsol+$U$+$V$ with the lattice geometry optimized using PBEsol (PBEsol+$U$+$V$@PBEsol). More specifically, even though the peaks~2 and 3 are still more red-shifted compared to experiments than in the PBEsol@PBEsol case, but now the peaks~5, 6, and 7 are in remarkable agreement with the experimental peaks. Therefore, while some peaks are more accurately described using PBEsol@PBEsol, others are better described using PBEsol+$U$+$V$@PBEsol, and hence none of these approximations provides the best description of all peaks in the Raman spectrum simultaneously. In contrast, the Raman spectrum computed using PBEsol+$U$ on top of the PBEsol+$U$ lattice geometry (PBEsol+$U$@PBEsol+$U$) diverges dramatically from the experimental one. To ascertain the influence of the underlying lattice geometry on the resulting Raman spectrum, we conducted a PBEsol+$U$ Raman spectrum calculation employing the PBEsol lattice geometry. This resultant spectrum, denoted as PBEsol+$U$@PBEsol, still exhibits poor alignment with the experimental data. This infers that the deficiency in the Raman spectrum of PBEsol+$U$@PBEsol+$U$ is not predominantly attributed to the PBEsol+$U$ lattice geometry, which remains cubic instead of rhombohedral (as elaborated in Sec.~\ref{structural}). The primary rationale for the ineffectiveness of PBEsol+$U$ resides in the fact that even with the adoption of the PBEsol lattice geometry, the atomic positions regress to the high-symmetry cubic orientations during the PBEsol+$U$ relaxation, thereby influencing the electronic structure significantly due to the $+U$ correction. Consequently, the covalency of the Ti($3d$)--O($2p$) bonding is suppressed, leading to a drastic deterioration of the Raman spectrum.

Overall, we ascertain that PBEsol@PBEsol and PBEsol+$U$+$V$@PBEsol yield the most accurate description of the Raman spectrum for rhombohedral BTO. The inclusion of the $+U+V$ corrections tends to red-shift low-wave-number Raman peaks but yields an excellent description of the high-wave-number Raman peaks, provided the PBEsol geometry is used. In stark contrast, the application of the $+U$ correction alone proves to be highly detrimental, leading to a Raman spectrum that deviates dramatically from the experimental data.

\section{Conclusions}
\label{sec:conclusions}

We have presented a detailed first-principles investigation of the low-temperature rhombohedral phase of BTO, focusing on its structural, electronic, and vibrational properties using PBEsol, PBEsol+$U$, and PBEsol+$U$+$V$. The onsite $U$ and intersite $V$ Hubbard parameters were computed using density-functional perturbation theory in a basis of L\"owdin-orthogonalized atomic orbitals. Our findings reveal that the application of the Hubbard $U$ correction to Ti($3d$) states localizes them and suppresses the Ti($3d$)--O($2p$) hybridizations, thus favoring the stabilization of the cubic phase in agreement with previous studies~\cite{Tsunoda_bandgap}. The inclusion of intersite Hubbard $V$ interactions between the Ti($3d$) and O($2p$) states preserves covalency of the Ti-O bonds, ultimately leading to the stabilization of the rhombohedral phase in agreement with experimental observations.

The optimized geometry in PBEsol+$U$+$V$ is found to be slightly less accurate than in PBEsol, resulting in a detrimental impact on the lattice vibrational properties. The Born effective charges for  $Z^{*}_{\mathrm{Ti}}$ components and $Z^{*}_{\mathrm{O}_{||}}$ are found to be smaller by $7-13\%$ compared to PBEsol results. The PDOS in PBEsol+$U$+$V$ is qualitatively similar to PBEsol, while the PBEsol+$U$+$V$ band gap and dielectric constant show good agreement with experiments by surpassing the accuracy of PBEsol predictions. The zone-center phonon frequencies, representing the positions of Raman peaks, are found to be very sensitive to the underlying geometry. The PBEsol and PBEsol+$U$+$V$ functionals provide the Raman spectra in satisfactory agreement with the experimental one, provided that the PBEsol geometry is used. 

Conversely, PBEsol+$U$ tends to stabilize the cubic BTO at 0~K due to the suppression of the Ti($3d$)--O($2p$) hybridizations, resulting in various properties that are in poorer agreement with experiments compared to PBEsol+$U$+$V$. Strikingly, the Raman spectrum computed using PBEsol+$U$ differs dramatically from the experimental one. Hence, the application of the $+U$ correction alone is found to be highly detrimental in the rhombohedral BTO.

Therefore, our study has uncovered the crucial significance of intersite Hubbard interactions in preserving the covalent features present in the rhombohedral phase of BTO. These findings could potentially extend beyond BTO and have broader implications for other ABO$_3$ perovskites exhibiting similar properties. However, further investigations employing PBEsol+$U$+$V$ for other materials are needed to fully explore the potential of this approach. Thus, our study contributes to laying the foundation for future research in this direction, offering new insights into tailoring the properties of these materials. The predictive power of PBEsol+$U$+$V$ holds promise for breakthroughs in perovskite materials, unlocking numerous opportunities for advanced technologies and applications.

\section*{Acknowledgments}

We thank Lorenzo Monacelli, Atsushi Togo, and Eugene Roginskii for fruitful discussions. This research was supported by the NCCR MARVEL, a National Centre of Competence in Research, funded by the Swiss National Science Foundation (Grant No. 205602). Computer time was provided by the Swiss National Supercomputing Centre (CSCS) under project No.~s1073, the Centre for High Performance Computing (CHPC), South Africa, and the supercomputer Lise and Emmy at NHR@ZIB and NHR@G\"{o}ttingen as part of the NHR infrastructure under the project \texttt{hbc00053} and \texttt{hbi00059}. The authors gratefully acknowledge support from the Deutsche Forschungsgemeinschaft (DFG) under Germany’s Excellence Strategy (EXC 2077, No. 390741603, University Allowance, University of Bremen) and Lucio Colombi Ciacchi, the host of the “U Bremen Excellence Chair Program", and the Abdus Salam International Centre for Theoretical Physics.


%

\end{document}